\documentclass[12pt]{article}

\usepackage{amssymb}
\usepackage{color}
\usepackage{graphicx,psfrag}
\usepackage{amsmath}
\usepackage{hyperref}

\def\be{\begin{equation}}
\def\ee{\end{equation}}
\def\bea{\begin{eqnarray}}
\def\eea{\end{eqnarray}}

\begin{document}

\begin{titlepage}

\setcounter{page}{1} \baselineskip=15.5pt \thispagestyle{empty}

\bigskip\
\begin{center}
{\Large \bf  Universality in D-brane Inflation}
\vskip 5pt
\vskip 15pt
\end{center}
\vspace{0.5cm}
\begin{center}
{
Nishant Agarwal,${}^1$ Rachel Bean,${}^1$ Liam McAllister,${}^2$ and Gang Xu${}^{2,3}$}
\end{center}

\vspace{0.1cm}

\begin{center}
\vskip 4pt
\textsl{${1}$  Department of Astronomy, Cornell University,
Ithaca, NY 14853}
\vskip 4pt
\textsl{${2}$  Department of Physics, Cornell University,
Ithaca, NY 14853}

\vskip 4pt
\textsl{${3}$  Institute for Advanced Study, Hong Kong University of Science and
Technology,}\\
\textsl{Clear Water Bay, Kowloon, Hong Kong}
\end{center} 
{\small  \noindent  \\[0.2cm]
\noindent
We study the six-field dynamics of D3-brane inflation for a general scalar potential on the conifold, finding simple, universal behavior.
We numerically evolve the equations of motion for an ensemble of more than $7 \cdot 10^7$ realizations, drawing the coefficients in the scalar potential from statistical distributions whose detailed properties have demonstrably small effects on our results.
When prolonged inflation occurs, it has a characteristic form: the D3-brane initially moves rapidly in the angular directions, spirals down to an inflection point in the potential, and settles into single-field inflation.
The probability of $N_{e}$ e-folds of inflation is a power law, $P(N_{e})\propto N_{e}^{-3}$, and we 
derive 
the same exponent from a simple analytical model.
The success of inflation is  relatively insensitive to the initial conditions: we find attractor behavior in the angular directions, and the D3-brane can begin far above the inflection point without overshooting.
In favorable regions of the parameter space, models yielding 60 e-folds of expansion arise approximately once in $10^3$ trials.  Realizations that are effectively single-field and give rise to a primordial spectrum of fluctuations consistent with WMAP, for which at least 120 e-folds are required, arise approximately once in $10^5$ trials.
The emergence of robust predictions from a six-field potential with hundreds of terms invites an analytic approach to multifield inflation.
}

\vspace{0.3cm}

\vspace{0.6cm}

\vfil
\begin{flushleft}
\small \today
\end{flushleft}
\end{titlepage}

\newpage
\tableofcontents
\newpage

\section{Introduction}

Inflation \cite{Guth:1980zm,Linde:1981mu,Albrecht:1982wi} provides a compelling explanation for the large-scale homogeneity of the universe and for the observed spectrum of cosmic microwave background (CMB) anisotropies.  However, in a large fraction of the multitude of inflationary models --- e.g., in most small-field models --- the success and the predictions of inflation are sensitive to small changes in the inflaton Lagrangian and initial conditions.
Without a priori measures on the space of scalar field Lagrangians and on the corresponding phase space, it is difficult to test a given model of inflation.

In this paper we find robust predictions in a surprising place: warped D-brane inflation \cite{Kachru:2003sx},
a well-studied scenario for inflation in string theory in which six (or more) dynamical fields are governed by a scalar potential with hundreds of terms.  We show that the collective effect of many terms in the potential is accurately described by a simple and predictive phenomenological model.  The essential idea behind our approach is that in an inflationary model whose potential involves the sum of many terms depending on multiple fields, one can expect a degree of emergent simplicity, which may be thought of as central limit behavior.

Our primary method is a comprehensive Monte Carlo analysis.  Recent results \cite{Baumann:2010sx} provide the structure of the scalar potential in warped D-brane inflation, i.e.\ a list of all possible terms in the potential, with undetermined, model-dependent coefficients.  A realization of warped D-brane inflation then consists of a choice of coefficients together with a choice of initial conditions.  We construct an ensemble of
realizations, drawing the coefficients from a range of statistical distributions and truncating the potential to contain $27$, $237$, and $334$ independent terms, corresponding to contributions from Planck-suppressed operators with maximum dimensions of $6$, $7$, and $\sqrt{28}-3/2 \approx 7.79$, respectively.
We then numerically evolve the equations of motion for the homogeneous background and identify robust observables that have demonstrably weak dependence on the statistical distribution, on the degree of truncation, and on the initial data.  In particular, we find that the probability of $N_{e}$ e-folds of inflation is a power law, $P(N_{e})\propto N_{e}^{-3}$, and we present a very simple analytical model of inflection point inflation that reproduces this exponent.

To study the primordial perturbations, we focus on
the subset of realizations in which the dynamics during the final 60 e-folds is that of single-field slow roll inflation.  (In the remaining realizations, multifield effects can be significant, and a dedicated analysis is required.)  For these cases, we find that primordial perturbations consistent with WMAP7 \cite{Larson:2010gs} constraints on the scalar spectral index, $n_{s}$, are possible only in realizations yielding $N_{e} \gtrsim 120$ e-folds.  In favorable regions of the parameter space, a universe consistent with observations arises  approximately once in $10^5$ trials.

The plan of this paper is as follows. In \S\ref{review} we recall the setup of warped D-brane inflation, and in \S\ref{method} we explain how we construct and study an ensemble of realizations.  Results for the homogeneous background evolution appear in \S\ref{background}, while the perturbations are studied in \S\ref{experiment}.   We conclude in \S\ref{conclusions}.  Appendix \ref{scalar potential} summarizes the structure of the inflaton potential, following \cite{Baumann:2010sx}.

\section{Review of Warped D-brane Inflation} \label{review}

In the simplest models of warped D-brane inflation,\footnote{See \cite{Dvali:1998pa}, \cite{Dvali:2001fw}, \cite{Burgess:2001fx} for foundational work on brane inflation.} the inflaton field $\phi$ is identified as the separation between a D3-brane and an anti-D3-brane along the radial direction of a warped throat region of a flux compactification \cite{Kachru:2003sx}.  (We will have much to say about more complicated inflationary trajectories that involve angular motion.)
The D3-brane potential receives a rich array of contributions, from the Coulomb interaction of the brane-antibrane pair, from the coupling to four-dimensional scalar curvature, and from nonperturbative effects that stabilize the K\"ahler moduli of the compactification.  The curvature coupling yields a significant inflaton mass,
and in the absence of any comparable contributions, the slow roll parameter $\eta$ obeys $\eta \approx 2/3$  \cite{Kachru:2003sx}, which is inconsistent with prolonged inflation.

The moduli-stabilizing potential does generically make significant contributions to the inflaton potential, and many authors have taken the attitude that within the vast space of string vacua, in some fraction the moduli potential will by chance provide an approximate cancellation of the inflaton mass, so that $\eta \ll 1$.  To do better, one needs to know the form of the moduli potential.  The nonperturbative superpotential was computed in \cite{Baumann:2006th} for a special class of configurations in which a stack of D7-branes falls inside the throat region.  For this case, Refs.~\cite{Burgess:2006cb,Baumann:2007np,Krause:2007jk,Baumann:2007ah} studied the possibility of inflation, and found that fine-tuned inflation, at an approximate inflection point, is indeed possible \cite{Baumann:2007np,Krause:2007jk,Baumann:2007ah}.

This situation is unsatisfactory in several ways. First, the restriction to compactifications in which D7-branes enter the throat is artificial, and serves to enhance the role of known terms in the inflaton potential (those arising from interactions with the nearby D7-branes) over more general contributions from the bulk of the compactification.  Second, the analyses of \cite{Burgess:2006cb,Baumann:2007np,Krause:2007jk,Baumann:2007ah,Panda:2007ie,Ali:2008ij,Chen:2008ai,Ali:2010jx} treated special cases in which the D3-brane tracked a minimum along some or all of the angular directions of the conifold, sharply reducing the dimensionality of the system, but there is no reason to believe that this situation is generic.  Therefore, although these works do provide consistent treatments of inflation in special configurations, an analysis that studies the full six-dimensional dynamics in a general potential is strongly motivated.\footnote{See \cite{Chen:2008ada,Chen:2010qz} for detailed studies of multifield effects at the end of D-brane inflation in the framework of \cite{Baumann:2007ah}, and \cite{Hoi:2008gc} for a systematic exploration of the likelihood of inflation in this context.}

The results of \cite{Baumann:2008kq, Baumann:2010sx} provide the necessary information about the D3-brane potential.  As explained in detail in \cite{Baumann:2010sx}, the most general potential for a D3-brane on the conifold corresponds to a general supergravity solution in a particular perturbation expansion around the Klebanov-Strassler solution.  The most significant terms in this potential arise from supergravity modes corresponding to the most relevant operators in the dual CFT, and by consulting the known spectrum of Kaluza-Klein modes, one can write down the leading terms in the inflaton potential, up to  undetermined Wilson coefficients.  The physical picture is that effects in the bulk of the compactification, e.g. gaugino condensation on D7-branes, or distant supersymmetry breaking, distort the upper reaches of the throat, leading to perturbations of the solution near the location of the D3-brane.

To describe the D3-brane action, we begin with the background geometry, which is a finite region of the warped deformed conifold.  Working far above the tip and ignoring logarithmic corrections to the warp factor, the line element is
\begin{equation}
ds^2 = \left(\frac{R}{r}\right)^2 g_{ij}dy^{i}dy^{j} = \left(\frac{R}{r}\right)^2 \Bigl(dr^2 +r^2 ds^2_{T^{1,1}}\Bigr)\,,
\end{equation} where $r$ is the radial direction of the cone, and the base space $T^{1,1}$ is parameterized by five angles,  $0\le \theta_1,\theta_2 \le \pi$, $0 \le \phi_1,\phi_2 < 2\pi$, $0 \le \psi < 4\pi$, which we shall collectively denote by $\Psi$.  The radius $R$ is given by $R^{4} =\frac{27}{4}\pi g_{s} N  {\alpha^{\prime}}^2$, with $N \gg 1$ the D3-brane charge of the throat.
At  radial coordinate $r_{\rm UV} \approx R$, the throat smoothly attaches to the remainder of the compact space, which we refer to as the bulk.  On the other hand, the deformation is significant in the vicinity of the tip, at $r_0 \approx a_{0} R$, with $a_{0}$ denoting the warp factor at the tip.  We will perform our analysis in the region $a_0 R \ll r < r_{\rm UV}$, where the singular conifold approximation is applicable, and will work with a rescaled radial coordinate  $x\equiv \frac{r}{r_{\rm UV}} <1$.
Finally, because the D3-brane kinetic term is insensitive to warping at the two-derivative level (cf.\ \S\ref{dbi}), the metric on the inflaton field space is the {\it{unwarped}} metric $g_{ij}$.

The D3-brane potential is usefully divided into four parts,
\begin{equation} \label{thepotential}
 V(x,\Psi)=V_0+V_{\cal{C}}(x) + V_{{\cal{R}}}(x) + V_{\rm bulk}(x,\Psi)\,,
\end{equation} 
which we will discuss in turn.
First, the constant $V_0$ represents possible contributions from distant sources of supersymmetry breaking, e.g. in other throats.
Next, the Coulomb potential $V_{\cal{C}}$ between an anti-D3-brane at the bottom of the throat and the mobile D3-brane has the leading terms
\begin{equation}
 V_{\cal{C}}=D_{0}\left(1-\frac{27D_{0}}{64\pi^2 T_{3}^2  r_{\rm UV}^4}\frac{1}{x^{4}}\right)\,,
\end{equation}
where  $T_3$ is the D3-brane tension and $D_{0}=2a_0^4 T_3$.  Higher-multipole terms in the Coulomb potential depend on the angles $\Psi$, but are suppressed by additional powers of $a_0$ and may be neglected  in our analysis.

Upon defining the scale $\mu^4=(V_0+D_0)\left(\frac{T_{3}r_{\rm UV}^{2}}{M_{\rm pl}^{2}}\right)$,  the leading contribution from curvature, corresponding to a conformal coupling, may be written
\begin{equation}
 V_{{\cal{R}}}=  \frac{1}{3}\mu^4 x^2\,.
\end{equation}

Finally, the structure of the remaining terms has been obtained in \cite{Baumann:2010sx}:
\begin{equation} \label{thepotential2}
 V_{\rm bulk}(x,\Psi)= \mu^4 \sum_{LM} c_{LM} x^{\delta(L)} f_{LM}(\Psi)\,.
\end{equation}
Here $LM$ are multi-indices encoding the quantum numbers under the $SU(2) \times SU(2) \times U(1)$ isometries of $T^{1,1}$, the functions $f_{LM}(\Psi)$ are angular harmonics on $T^{1,1}$, and $c_{LM}$ are constant coefficients.
The exponents $\delta(L)$ have been computed in detail in \cite{Baumann:2010sx}, building on the computation of Kaluza-Klein masses in \cite{Ceresole}:
\begin{equation}
\delta=1,\, 3/2,\, 2,\,\sqrt{28}-3,\, 5/2,\,\sqrt{28}-5/2,\,3,\,\sqrt{28}-2,\,7/2,\,\sqrt{28}-3/2, \ldots
\end{equation}
From the viewpoint of the low-energy effective field theory, a term in the potential proportional to $x^{\delta(L)}$ arises from a Planck-suppressed operator with dimension $\Delta=\delta(L)+4$. In particular, the conformal coupling to curvature corresponds to an operator of dimension six, ${\cal O}_{6} =(V_0+D_0)\phi^2$.

Two technical remarks are in order.  First, for  simplicity of presentation we have included higher-order curvature contributions in the list of bulk terms, rather than in a separate category.  Second, perturbations of the unwarped metric $g_{ij}$ lead to terms in the D3-brane potential that were not analyzed in \cite{Baumann:2010sx}, but can be important in some circumstances.\footnote{We thank Sohang Gandhi for very helpful discussions of this point.}  We do not implement the angular structure of these terms in full detail, but we have verified that these contributions lead to negligible corrections to our results.

The coefficients $c_{LM}$  could be computed in principle in a specific realization in which  all details of the compactification are available, but in practice must be treated as unknown parameters.
Our approach is to assume that all possible terms are present, with coefficients $c_{LM}$ of comparable magnitude.  Specifically, we will draw the   $c_{LM}$ from a range of statistical distributions and then verify that the
(unknown) detailed statistical properties of the $c_{LM}$ are not important for the inflationary phenomenology, while the overall scale of the $c_{LM}$ does matter significantly.

We set the overall scale of the bulk contributions by noting that the moduli potential and the remainder of the potential are tied by the requirement that the cosmological constant should be small after brane-antibrane annihilation.  With our definition of $\mu$, the scaling arguments presented in Appendix A of \cite{Baumann:2008kq} suggest that in typical KKLT compactifications, $c_{LM} \sim {\cal{O}}(1)$.


Let us remark that the potential (\ref{thepotential}) is not the most general function  on the conifold: it is the most general D3-brane scalar potential on the conifold (within the fairly broad assumptions of \cite{Baumann:2010sx}.)  Many terms in (\ref{thepotential}) enjoy correlations that would be absent in a totally general function, and which arise here because certain physical sources, such as fluxes, contribute in correlated ways to different terms in the potential.  We defer a full description of the construction of the potential to Appendix \ref{scalar potential}.

Finally, we note that in compactifications preserving discrete symmetries that act nontrivially on the throat region, the structure of the D3-brane potential is altered by the exclusion of terms that are odd under the discrete symmetries \cite{Baumann:2008kq}. Exploring the phenomenology of the corresponding models is an interesting question that is beyond the scope of this work.\footnote{We thank Daniel Baumann for helpful discussions of this point.}

\section{Methodology} \label{method}
To characterize the dynamics of D3-brane inflation in a general potential, we perform a Monte Carlo analysis, numerically evolving more than $7 \cdot 10^7$ distinct realizations of the model.
In this section we explain our recipe for constructing an ensemble of realizations.   In \S\ref{evolution}, we obtain the
equations of motion and introduce the parameters required to specify the potential.  In \S\ref{parameters} we describe how we draw the coefficients in the potential from statistical  distributions, and in
in \S\ref{rotational} we indicate how we choose initial conditions.
\subsection{Setup}\label{evolution}

The inflaton field is characterized by one radial coordinate and five angular coordinates. At the two-derivative level (see \S\ref{dbi} for a discussion of DBI effects), the equations of motion for the homogeneous background are the Klein-Gordon equations obtained from the Lagrangian
\begin{equation}
	\mathcal{L} = a^{3}\left(\frac{1}{2}T_3 g_{ij}\dot{y}^{i}\dot{y}^{j} - V(y) \right)\,,
\end{equation}
where $a$ is the scale factor, along with the Friedmann and acceleration equations,
%
\bea
 3H^{2} & = & \frac{1}{2}T_3 g_{ij}\dot{y}^{i}\dot{y}^{j} + V(y)\,, \\
 \dot{H} & = & -\frac{1}{2}T_3 g_{ij}\dot{y}^{i}\dot{y}^{j}\,.
\eea
Here
$y^{i}$ denotes the six coordinates, dots indicate derivatives with respect to time, $H \equiv \dot{a}/a$, and $g_{ij}$ is the metric on the  inflaton field space, which is the conifold.

Three important microphysical parameters are the D3-brane tension, $T_3$; the length of the throat, $\phi_{\rm UV} \equiv r_{\rm UV} \sqrt{T_{3}}$; and the warp factor at the tip, $a_{0}$.  The combination $2 T_3 a_{0}^{4} \equiv D_{0}$ determines the overall scale of inflation, while  $\phi_{\rm UV}$ dictates the size of the field space.

The result of \cite{Baumann:2006cd} gives an upper bound on the inflaton field range,  $\phi_{\rm UV}<\frac{2 M_{\rm pl}}{\sqrt{N}}$, with $N \gg 1$ the D3-brane charge of the warped throat.  Consistent with this, we take $\phi_{\rm UV}=0.1$.  Working in units where $M_{\rm pl}^{-2}=8\pi G =1$ for the remainder, we set the D3-brane tension to be $T_3=10^{-2}$, and for our analysis of the background evolution, we take $a_0=10^{-3}$.
We have verified that changing these parameters does not substantially alter our results
for the homogeneous background. However, changing $T_3 a_{0}^{4}$ --- which we accomplish by changing $a_{0}$ --- does affect the scale of inflation, and hence the normalization of the scalar perturbations.  Therefore, in our study of the perturbations in \S\ref{experiment}, we scan over a range of values for $a_0$, focusing on values most likely to lead to a WMAP-normalized spectrum \cite{Larson:2010gs}.  This is a fine-tuning that we will not attempt to quantify, as there is no agreed-upon measure for $a_0$.


\subsection{Constructing an ensemble of potentials} \label{parameters}

In principle the D3-brane potential (\ref{thepotential}) has an infinite number of terms, 
but  for $x\equiv \frac{r}{r_{\rm UV}} < 1$ one can truncate (\ref{thepotential}) at some maximum exponent $\delta=\delta_{\rm max} \equiv \Delta_{\rm max} -4$.
Because of the critical role of the inflaton mass term, truncating to $\Delta_{\rm max}<6$ would fail to capture essential physical properties, so we must have  $\Delta_{\rm max} \geq 6$.  As $\Delta$ increases, the number of independent terms grows very rapidly, because there are 
many angular harmonics $f_{LM}$ for each $\Delta$.
Limited by computational power, we truncate the potential at $\Delta_{\rm max}= \sqrt{28}+5/2 \approx 7.8$.  We perform identical analyses for $\Delta_{\rm max}=6$, $7$, and $7.8$, corresponding  respectively to 27, 237 and 334 independent terms in the potential, in order to assess whether our results are sensitive to the cutoff.

Many studies of D-brane inflation treat the evolution of the radial position, the  volume of a particular four-cycle, and sometimes one angular coordinate, cf.\ e.g.\ \cite{Burgess:2006cb,Baumann:2007np,Krause:2007jk,Baumann:2007ah,Panda:2007ie,Ali:2008ij,Chen:2008ai,Ali:2010jx}, rather than the full multifield dynamics.  Although angular evolution in the framework of \cite{Baumann:2007ah} has been studied in detail in \cite{Chen:2008ada,Chen:2010qz}, the focus of these works was the onset of angular instabilities at the end of inflation. We will find that angular evolution before the onset of inflation also plays a critical role.

To understand how our results differ from treatments with fewer dynamical fields, we study the impact of stepwise increases in the number $N_f$ of evolving fields.   We artificially, but self-consistently, freeze $6-N_f$ of the angular fields by not imposing the corresponding equations of motion, creating realizations that depend on $N_f$ variables.   While these realizations have less physical meaning than the full potential, they provide some insight into the role of the angular fields. For simplicity we study $N_f=1$, $N_f=2$, and the full case $N_{f}=6$.

As we are not assuming that D7-branes wrapping a four-cycle descend into the throat region, we will not model the evolution of the K\"ahler moduli.  Although it would be very interesting (and challenging) 
to study the cosmological dynamics of K\"ahler moduli in the bulk, the universality found in the present analysis makes it  plausible that additional fields would have little effect on the inflationary phenomenology.

Turning now to the Wilson coefficients $c_{LM}$, we do not assume a specific compactification, but instead draw the $c_{LM}$  from a range of statistical distributions.
We define the root mean square (rms) size, $\langle c_{LM}^2 \rangle^{1/2} \equiv Q$, where the brackets denote the ensemble average, and by assumption\footnote{A strong trend in $Q$ as a function of $\Delta$ could change the relative importance of terms with large $\Delta$, and hence affect our conclusions about the robustness of the truncation to $\Delta \le \Delta_{\rm max}$.  We are not aware of a well-motivated proposal for such a trend, but it could be worthwhile to investigate this further.}  the rms size is independent of $L$ and $M$.
It is then convenient to write
\begin{equation}\label{qhat}
c_{LM} = Q \,\hat{c}_{LM}\,,
\end{equation} and draw the $\hat{c}_{LM}$ from some distribution ${\cal M}$ that has unit variance but is otherwise arbitrary.

The physical picture is that $Q$ depends on the distance to the nearest stack of D7-branes effecting K\"ahler moduli stabilization.  The estimates performed in \cite{Baumann:2008kq} indicate that $Q \sim {\cal{O}}(1)$ for
D7-branes in the upper region of the throat. We anticipate that as the nearest D7-branes are moved farther into the bulk, $Q$ will diminish to some extent, though we are not aware of a regime in which the bulk contributions are strictly negligible.

If the inflationary phenomenology depended  in detail on the nature of ${\cal M}$, e.g. if the success of inflation depended sensitively on the higher moments of ${\cal M}$, then no general predictions would be possible.  Let us clarify that dependence on the rms size $Q$ of the $c_{LM}$, corresponding to the typical size of the bulk contribution to the inflaton potential, is to be expected and is not problematic. Difficulty would arise if, for example,  two distributions with unit variance but with distinct skewness or kurtosis led to disparate predictions.

There are strong motivations for expecting  that some statistical properties of the potential will be  independent of ${\cal M}$.  For example, if a symmetric $N \times N$ matrix has its
entries drawn from some distribution with appropriately bounded moments, then in the large $N$ limit the statistical properties of the eigenvalues  are indistinguishable from those obtained from entries drawn from a Gaussian distribution with mean zero \cite{Bai}.
By experimenting with different distributions, we will identify observables which, like the eigenvalue distribution in random matrix theory, are robust against changes in the statistics of the inputs.
In practice, for much of our analysis we choose ${\cal M}$ to be a Gaussian distribution with mean zero,
and then carefully verify for a range of other distributions that our results receive negligible corrections.

\subsection{Initial conditions}\label{rotational}

The phase space of initial conditions for a D3-brane in the conifold is 12-dimensional: six dimensions for the initial positions $x_{0},\Psi_{0}$ and another six dimensions for the initial velocities $\dot{x}_{0}, \dot{\Psi}_{0}$. A grid-based scan across the full 12-dimensional space would be very computationally intensive even with only a few points along each dimension. Fortunately, five of the six dimensions are angular coordinates on the coset space $T^{1,1}$, which has a large isometry group, $SU(2) \times SU(2) \times U(1)$. These isometries can be used to reduce the dimensionality of the initial phase space, in the following way.
A generic configuration of sources in the compact space will break the isometry group completely, but in a large ensemble of realizations, we expect  that there are no preferred regions on $T^{1,1}$: the ensemble averages should respect the isometries even though any individual realization breaks the isometries.
Thus, without loss of generality we may pick a fixed point  $\Psi_{0}$ on $T^{1,1}$ for the initial position.  For numerical purposes it is convenient to begin away from the coordinate singularities, so we choose $\Psi_{0}$ to be $\theta_1=\theta_2=\phi_1=\phi_2=\psi=1.0$.

The initial angular velocities $\dot{\Psi}_{0}$ are slightly more complicated.\footnote{We are grateful to Raphael Flauger for helpful discussions of this point.}  To describe a general angular velocity, it suffices to specify the magnitude of the velocity in each $S^2$ and in the fiber $S^{1}$.  For simplicity we focus on velocity in the fiber, $\dot{\psi}$, and take the remaining components of the initial velocity to vanish.
We expect, and find, similar results for initial velocities in either $S^2$, but we postpone a complete scan of the phase space to future work.

We are left with a three-dimensional space of initial configurations spanned by the radial position $x_{0}$, the radial velocity $\dot{x}_{0}$, and the angular velocity $\dot{\Psi}_{0}=\dot{\psi}_{0}$.  Of course, our evolution occurs in the full 12-dimensional phase space: the simplification applies only to the initial conditions.  For a portion of our Monte Carlo analysis, we set  $\dot{x}_{0}=\dot{\psi}_{0}=0$, so that the D3-brane begins at rest.  In \S\ref{iniv} we describe the effect of nonvanishing initial velocities.

\subsection{Parameters summarized}

To summarize, we fix $T_3=10^{-2}$, $a_0=10^{-3}$, and $\phi_{\rm UV}=0.1$  for our analysis of the  background evolution.
We truncate the D3-brane potential to include contributions from operators with maximum dimension $\Delta_{\rm{max}}=6$, $7$, and $7.8$, and we take $N_f=1,2,6$ of the D3-brane coordinates to be dynamical fields. The coefficients $c_{LM}$ have rms size $Q$, and the rescaled quantities $\hat{c}_{LM} = c_{LM}/Q$ are drawn from a distribution ${\cal M}$ that has unit variance.
We begin at $x=x_{0}\equiv 0.9$, $\Psi=\Psi_{0} \equiv \{\theta_1=\theta_2=\phi_1=\phi_2=\psi=1.0\}$, with arbitrary radial velocity $\dot{x}_{0}$, arbitrary angular velocity $\dot{\psi}_{0}$ in the $\psi$ direction, and all other angular velocities vanishing.
We would now like to understand how the observables depend on the input parameters $Q$, $\Delta_{\rm{max}}$, $N_f$, and ${\cal M}$, and on the initial data $x_{0},\dot{x}_{0},\dot{\psi}_{0}$.

\begin{figure}[!h]
  \begin{center}
    \includegraphics[width=3.0in,angle=0]{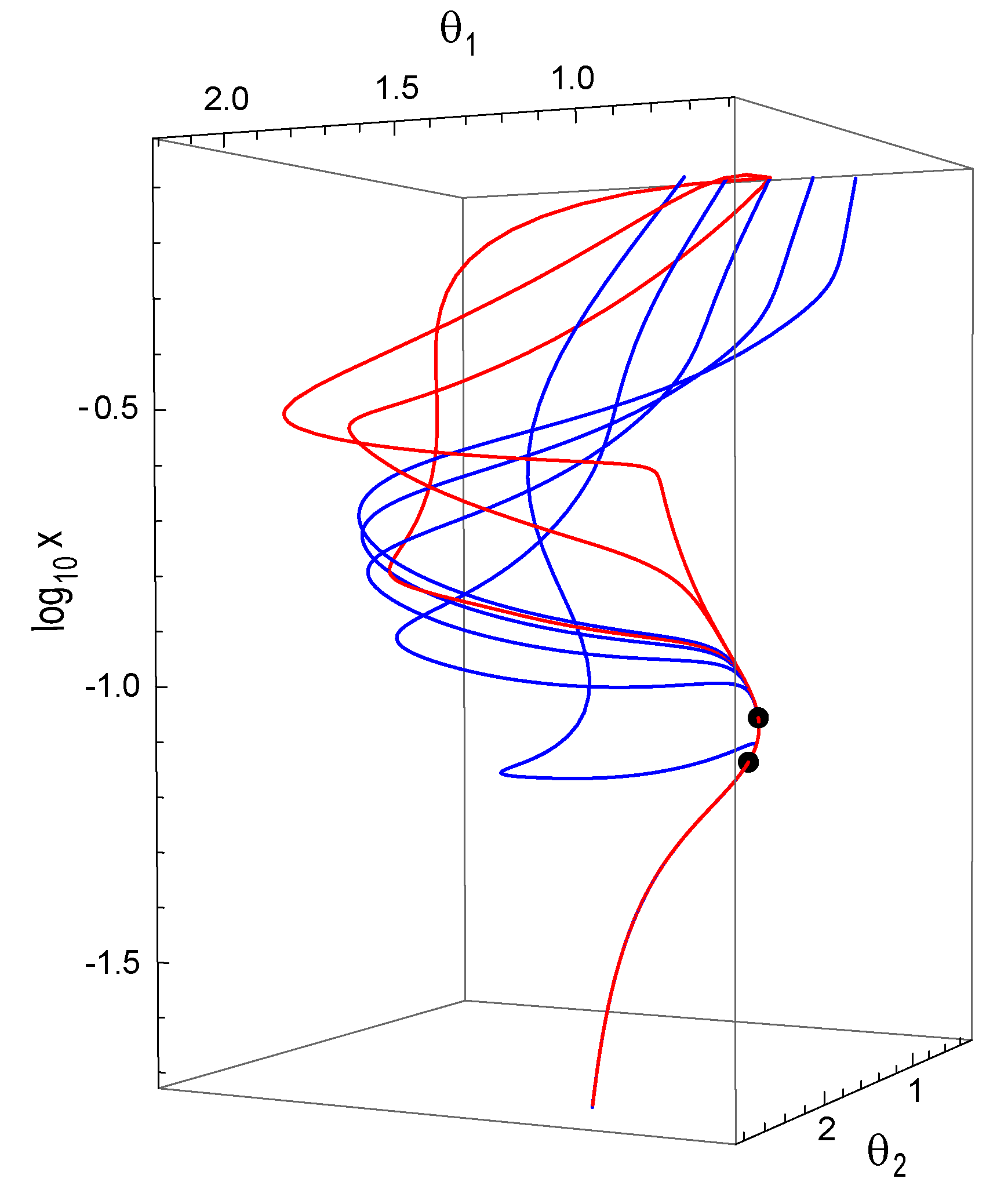}
    \caption{Examples of downward-spiraling trajectories for a particular realization of the potential. The black dots mark 60 and 120 e-folds before the end of inflation (7 of the 8 curves shown achieve $N_e>120$); inflation occurs along  an inflection point that is not necessarily parallel to the radial direction.  Red curves have nonvanishing initial angular velocities $\dot{\Psi}_{0}$, while blue curves have $\dot{\Psi}_{0}=0$.}
    \label{bouquet}
  \end{center}
\end{figure}

\section{Results for the Homogeneous Background}
\label{background}

As a first step, we study the evolution of the homogeneous background.   In \S\ref{probability}, we show that for fixed initial conditions, the probability of $N_e$ e-folds of inflation is a power law, and we show that the exponent is robust against changes in the input parameters $\Delta_{\rm max}$, $N_f$, and ${\cal{M}}$.
In \S\ref{analyticsection} we present a simple analytic model that reproduces this power law.  We study the effect of varying the initial conditions in \S\ref{initial}, and we discuss DBI inflation in \S\ref{dbi}.

\subsection{The probability of inflation}
\label{probability}


We find it useful to divide possible trajectories into three classes.  The D3-brane can be ejected  from the throat, reaching $x>1$ and leaving the domain of validity of our analysis; it can become trapped in a local or global minimum of the potential; and it can reach the bottom\footnote{In practice, we define the bottom of the throat to be at $x = 20 a_0$ in order to remain well above the region where the throat rounds off and  the singular conifold approximation fails.}
of the throat, triggering the hybrid exit and reheating, after a certain number of e-folds of inflation.

A central question is what fraction of realizations solve the horizon problem by producing $N_{e} \ge 60$ e-folds of inflation, and then plausibly transition to the hot Big Bang.  We will not model reheating in detail, but we will insist that only e-folds of inflation {\it{that precede a hybrid exit}} are counted towards $N_{e}$.   That is, false vacuum inflation in a metastable minimum, or slow roll inflation preceded by ejection and unknown dynamics in the bulk, do not contribute to  $N_{e}$.

Specifically, for $k$ distinct trials we define
\begin{equation}
P(N_e>60) = \frac{\#(N_e>60)}{\#(N_e>60)+\#(N_e\le60)+\#(\rm{ejected})+\#(\rm{trapped})} \,,
\end{equation}
i.e.\ trials leading to ejection or trapping are included in the denominator, so that $P(N_e>60)$ reflects the probability of $N_{e} > 60$ e-folds of inflation preceding a hybrid exit
in a general realization.

\begin{figure*}[!h]
  \begin{center}
  \mbox{
    {\includegraphics[width=3.in,angle=0]{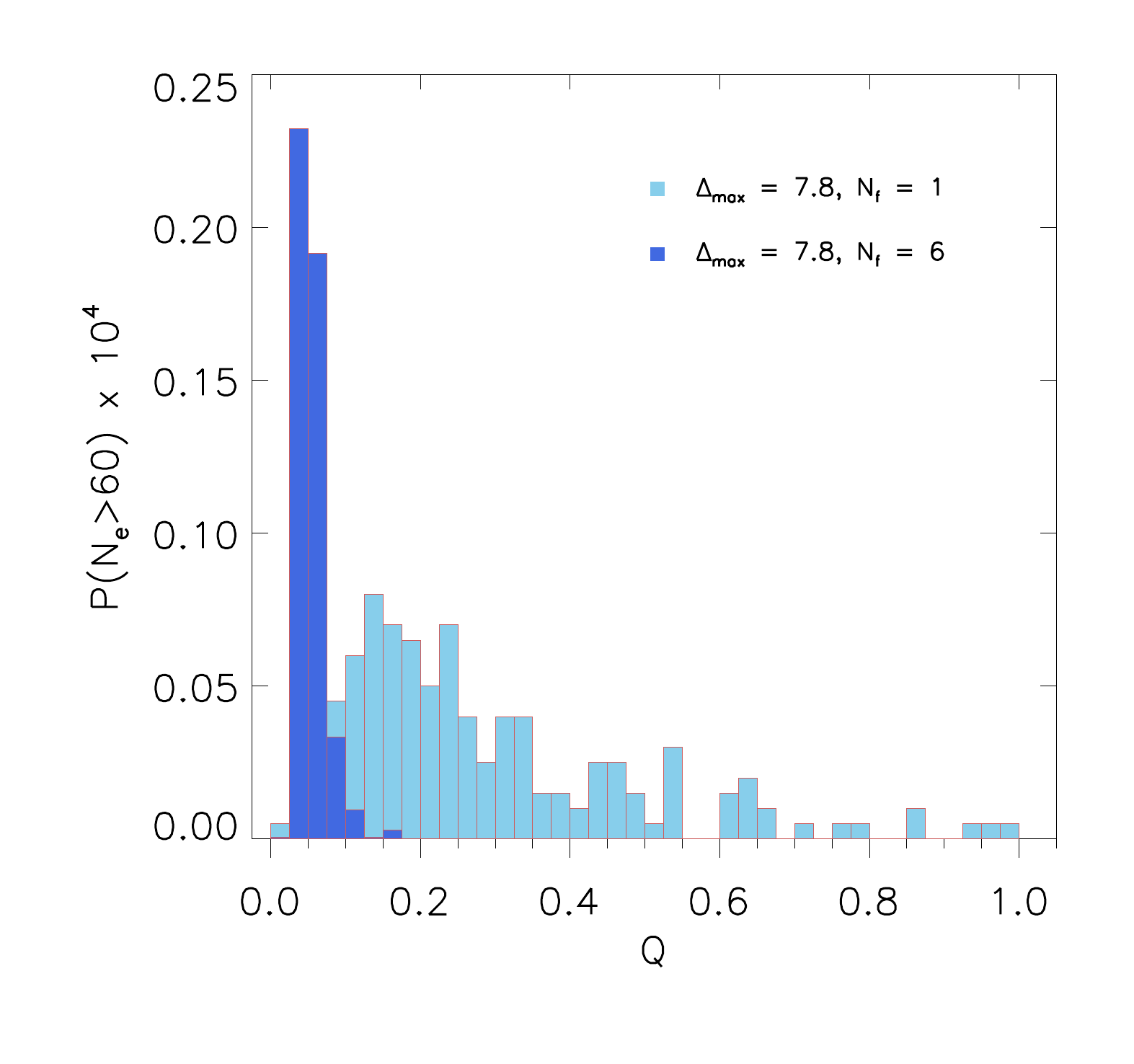}}\quad
{        \includegraphics[width=3.in,angle=0]{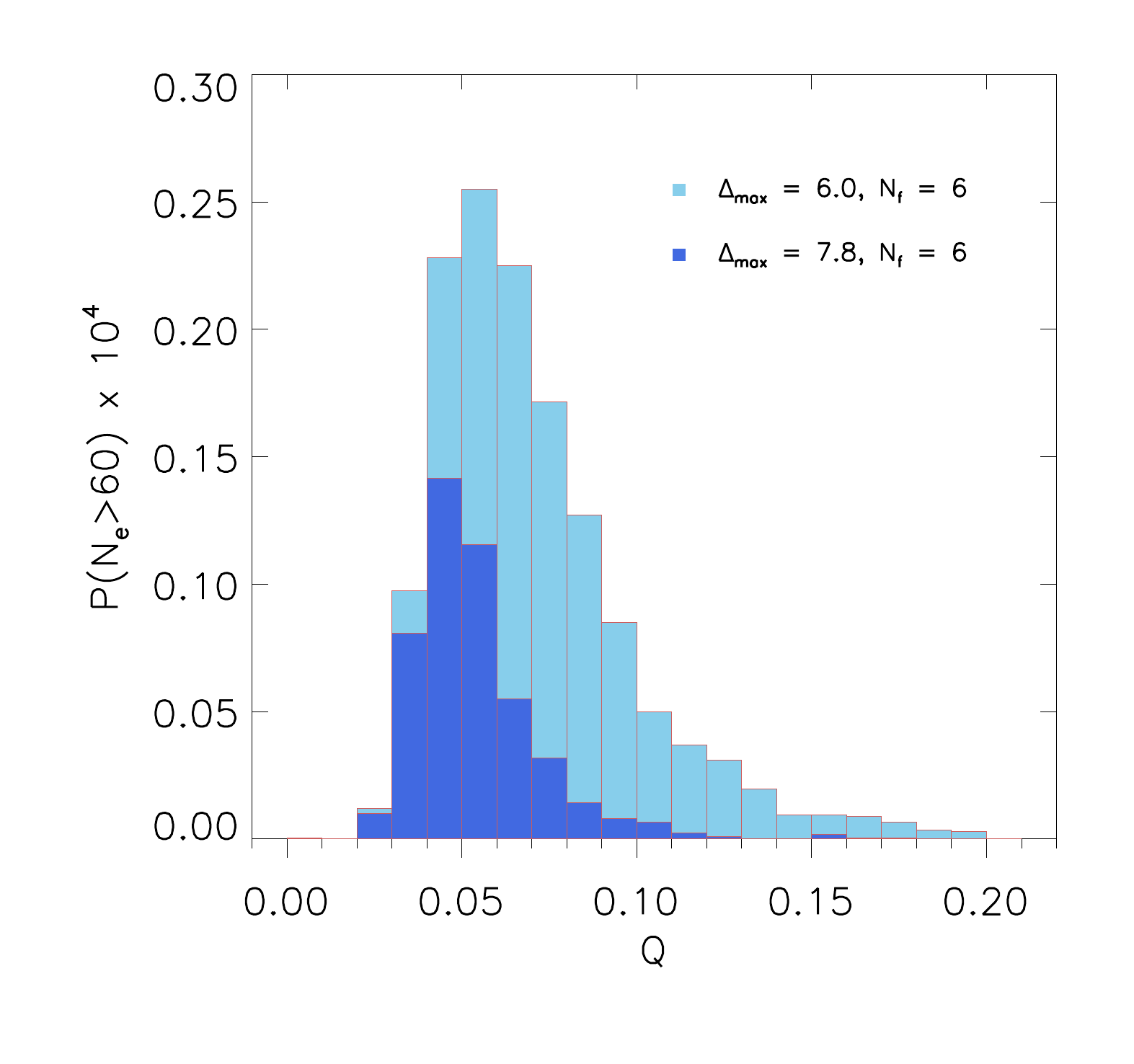}}
}
    \caption{The rms value, $Q$, of the coefficients $c_{LM}$ has a significant role in determining whether inflation can occur.
    [Left panel]  The success probability $P(N_e>60)$ for two different numbers of fields, $N_f=1$ and $N_f=6$, with $\Delta_{\rm max}=7.8$.  [Right panel] The success probability for two different degrees of truncation, $\Delta_{\rm max}=6$ and $\Delta_{\rm max}=7.8$, with $N_f=6$.}
    \label{fig:Q_histogram}
  \end{center}
\end{figure*}

We now examine how the `success probability' $P(N_e>60)$ depends on the input parameters $Q$, $\Delta_{\rm max}$, $N_f$, and ${\cal{M}}$.  First, Figure \ref{fig:Q_histogram} shows that $P(N_e>60)$ depends strongly on
$Q$, and
the optimal value of $Q$ depends on $\Delta_{\rm max}$ and on $N_f$.
When all six fields are dynamical ($N_f=6$), the probability of inflation is optimized for $Q\sim 0.04$, while for $N_f=1$, $Q\sim1$ can yield sufficient inflation.

To understand this result, we recall that in the presence of a single harmonic contribution to the inflaton potential, after minimization of the angular potential, the radial potential is expulsive \cite{Baumann:2008kq}.  More generally, the bulk contributions to the potential provide the only possibility of counterbalancing the Coulomb and curvature contributions, which both draw the D3-brane towards the tip. For $Q=0$, the Coulomb and curvature contributions are not counterbalanced, and the D3-brane falls quickly towards the tip without driving inflation. For $Q \sim 1$, a single harmonic contribution term could marginally balance the inward force; the net effect of 334 such terms then plausibly leads to rapid expulsion from the throat.  This result is consistent with our finding that the optimal value of $Q$ diminishes as the number of terms in the potential increases,  as shown in Figure \ref{fig:Q_histogram}.

\begin{figure}[!h]
  \begin{center}
          \includegraphics[width=3.in,angle=0]{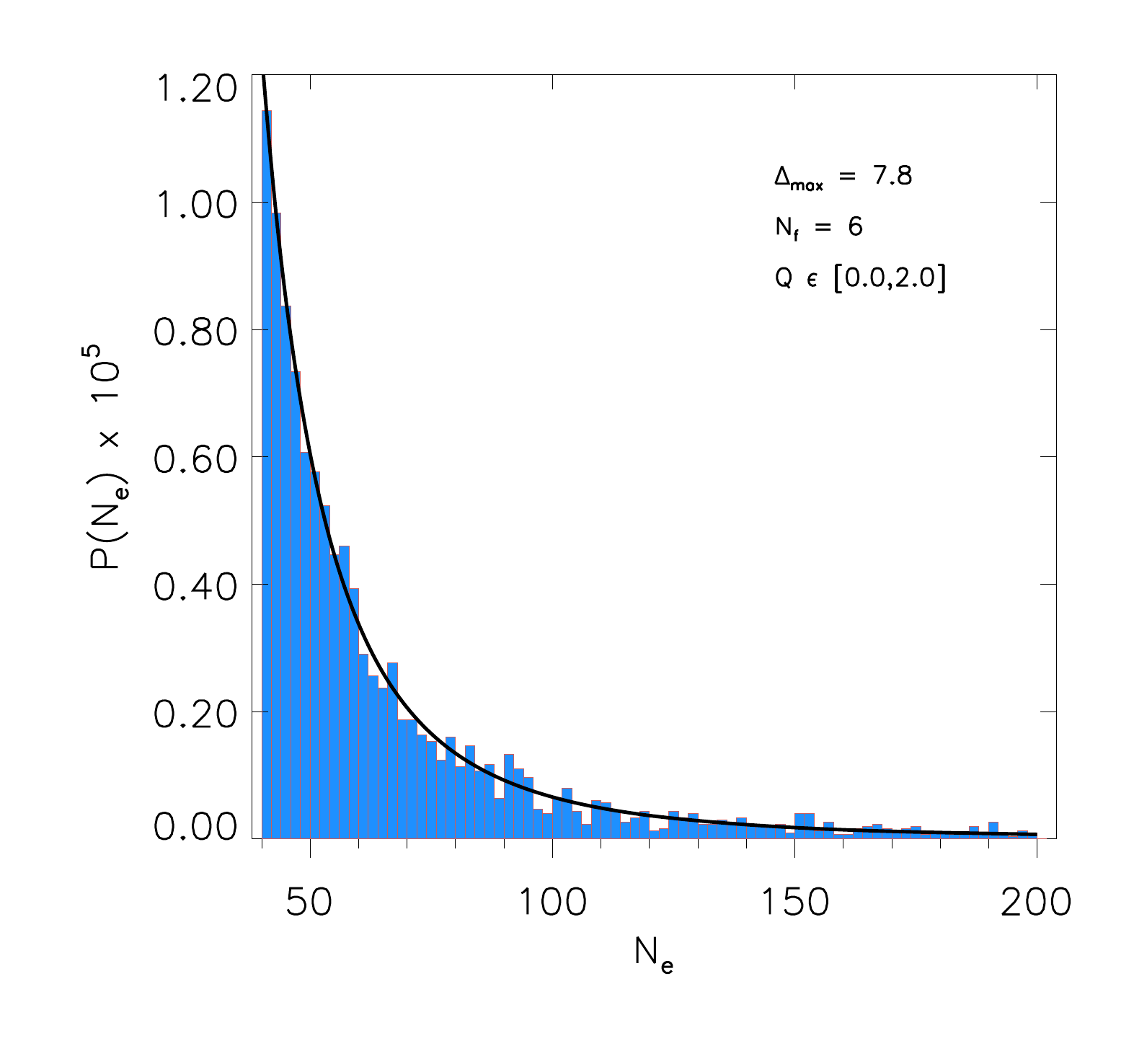}
          \includegraphics[width=3.in,angle=0]{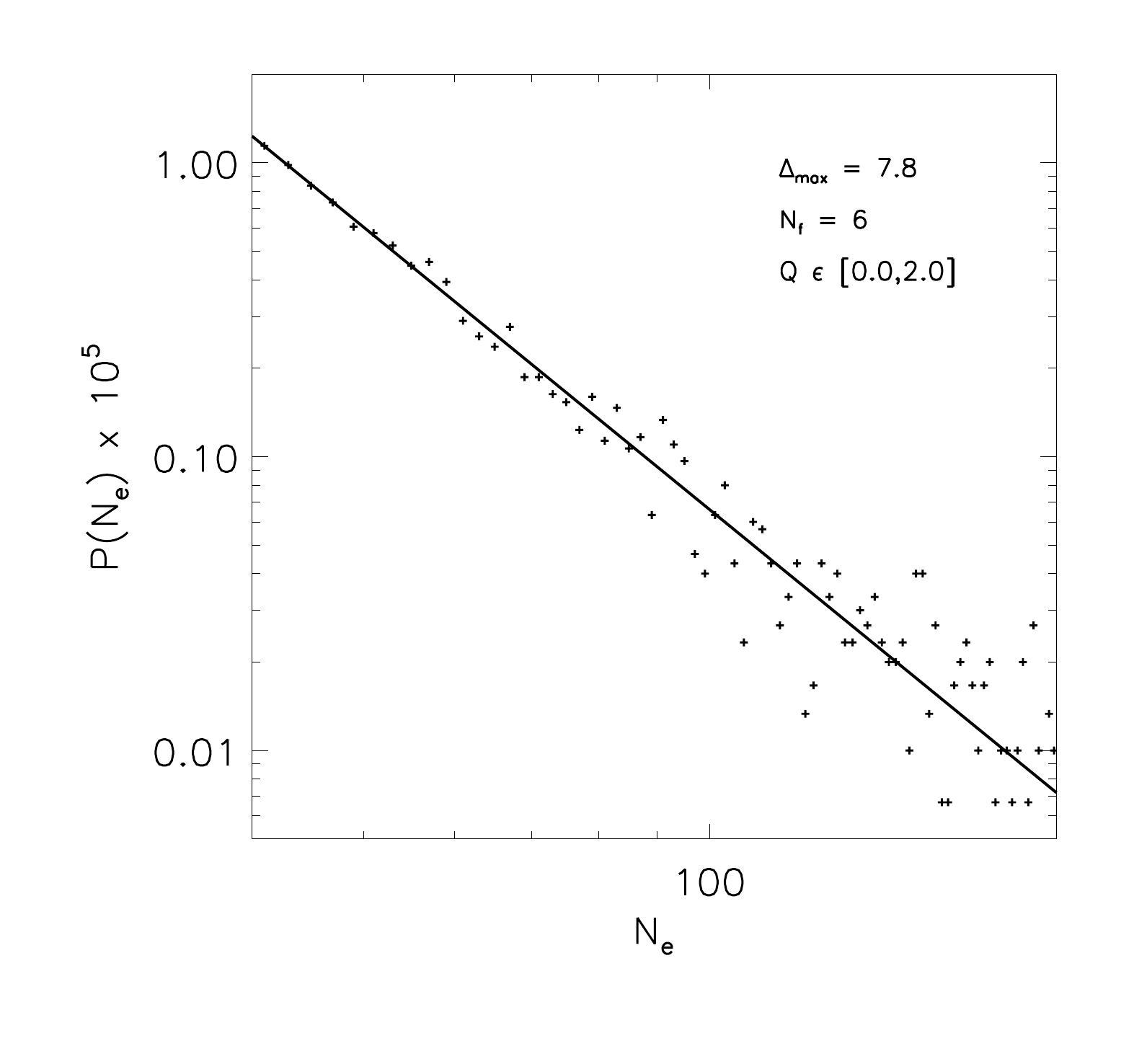}
    \caption{The likelihood of $N_{e}$ e-folds of inflation as a function of $N_{e}$, for $\Delta_{\rm max} = 7.8$ and $N_{f} = 6$.
    We find $P(N_e)\propto N_e^{-\alpha}$, with $\alpha=3.22\pm 0.07$ at the 68\% confidence level. The left panel shows the power law fit to the histogram and the right panel shows the same fit on a log-log plot.}
    \label{fig:gamma_histogram}
\end{center}
\end{figure}

In Figure \ref{fig:gamma_histogram} we display a histogram of Monte Carlo trials that give more than 40 e-folds of inflation for $Q \in [0,2.0]$ with $\Delta_{\rm max}=7.8$ and all six fields evolving ($N_{f} = 6$). We find that we can characterize the probability of inflation, for scenarios yielding $N_{e} \gg 10$ e-folds, by a function $P(N_{e}) =  {\cal{A}}\, (N_{e}/60)^{-\alpha}$. On fitting the data in Figure \ref{fig:gamma_histogram}, we find that $\alpha = 3.22 \pm 0.07$  and  ${\cal{A}} = 1.7 \cdot 10^{-6}$.

\begin{table}[t!]
\begin{center}
\begin{tabular}{|c|c|c|c|c|c|c|c|}
\hline
$N_f$ & $\Delta_{\rm max}$ & ${\mathcal A}$ & $\alpha$ & $P(N_{e}>60)$ & $P(N_{e}>120)$ & $P({\rm ej})$ & $P({\rm min})$
\\
\hline
6 & 7.8  &  $1.7 \cdot 10^{-6}$  &  $3.22 \pm 0.07$  &  $4.6 \cdot 10^{-5}$  &  $1.1 \cdot 10^{-5}$ & $0.97$ & $3.9 \cdot 10^{-3}$
\\
6 & 7    &  $3.2 \cdot 10^{-6}$  &  $3.19 \pm 0.10$  &  $9.0 \cdot 10^{-5}$  &  $2.0 \cdot 10^{-5}$ & $0.96$ & $8.8 \cdot 10^{-3}$
\\
6 & 6    &  $4.3 \cdot 10^{-6}$   &  $2.85 \pm 0.14$  &  $1.3 \cdot 10^{-4}$  &  $3.3 \cdot 10^{-5}$ & $0.92$ & $5.4 \cdot 10^{-2}$
\\
\hline
2 & 7.8  &  $3.0 \cdot 10^{-5}$  &  $3.30 \pm 0.25$  &  $9.6 \cdot 10^{-5}$  &  $2.4 \cdot 10^{-5}$ & $0.77$ & $6.7 \cdot 10^{-2}$
\\
2 & 7    &  $5.6 \cdot 10^{-5}$  &  $2.97 \pm 0.17$  &  $1.8 \cdot 10^{-4}$  &  $4.9 \cdot 10^{-5}$ & $0.66$ & $1.2 \cdot 10^{-1}$
\\
2 & 6    &  $1.5 \cdot 10^{-5}$  &  $2.99 \pm 0.12$  &  $4.7 \cdot 10^{-4}$  &  $1.3 \cdot 10^{-4}$ & $0.51$& $1.9 \cdot 10^{-1}$
\\
\hline
1 & 7.8  &  $2.7 \cdot 10^{-6}$  &  $2.83 \pm 0.22$  &  $8.8 \cdot 10^{-5}$  &  $2.6 \cdot 10^{-5}$ & $0.36$ & $5.1 \cdot 10^{-2}$
\\
1 & 7    &  $4.3 \cdot 10^{-6}$  &  $3.03 \pm 0.21$  &  $1.2 \cdot 10^{-4}$  &  $2.9 \cdot 10^{-5}$ & $0.29$ & $6.1 \cdot 10^{-2}$
\\
1 & 6    &  $1.3 \cdot 10^{-5}$   &  $2.86 \pm 0.15$  &  $4.2 \cdot 10^{-4}$  &  $1.1 \cdot 10^{-4}$ &$0.17$ &$7.4 \cdot 10^{-2}$
\\
\hline
\end{tabular}
\caption{Summary of the power law fits to the success probabilities for various scenarios, with $Q$ varied over the range $[0.0,2.0]$. We fit the Monte Carlo data for $N_e>40$ to a power law of the form $P(N_e) = {\cal{A}}\,(N_e/60)^{-\alpha}$.  The two final columns indicate the probabilities that the D3-brane will be ejected from the throat or trapped in a minimum, respectively.
}
\label{table:gamma}
\end{center}
\end{table}

In Table \ref{table:gamma} we summarize the power law fits to the probability of inflation as one considers different numbers of fields, $N_f$, and different truncations of the potential,  $\Delta_{\rm max}$, assuming that ${\cal M}$ is Gaussian, and taking zero initial angular and radial velocities at $x=0.9$ and fixed angular position $\Psi_{0}$.

The probability of obtaining 60 e-folds of inflation does not change dramatically if one truncates the potential at  $\Delta_{\rm max} = 6.0$ or $\Delta_{\rm max} = 7.8$, so our results appear insensitive to the precise placement of the truncation.  As the number of fields, $N_f$, increases, the range of $Q$ yielding inflation becomes restricted, but we also find that the probability of achieving inflation within this $Q$ range increases, cf.\ Figure \ref{fig:Q_histogram}.  In fact, the power law fit of the success probability remains fairly consistent as $N_f$ is varied, provided that one marginalizes over $Q$.

Although we have seen that $P(N_e>60)$ is highly sensitive to the value of $Q$, we find that $P(N_e>60)$ has negligible dependence on the shape of the distribution ${\cal M}$ from which the $\hat{c}_{LM}$  are drawn.
Specifically, we have obtained power law fits of the success probability for ensembles in which ${\cal M}$ is a Gaussian, shifted Gaussian, triangular, or uniform distribution.
As shown in Table \ref{table:measure}, we find negligible changes in ${\cal A}$ and $\alpha$.
We expect that there exist pathological distributions, e.g.\ with rapidly growing higher moments, that could change our findings, but we are not aware of a microphysical argument for such a distribution.

\begin{table}[t!]
\begin{center}
\begin{tabular}{|c|c|c|c|c|c|c|c|}
\hline
${\cal M}$ & $\Delta_{\rm max}$ & ${\mathcal A}$ & $\alpha$ & $P(N_{e}>60)$ & $P(N_{e}>120)$
\\
\hline
Gaussian& 7.8  &  $1.6 \cdot 10^{-6}$  &  $3.21 \pm 0.14$  &  $4.6 \cdot 10^{-5}$  &  $1.1 \cdot 10^{-5}$
\\
shifted & 7.8    &  $1.7 \cdot 10^{-6}$  &  $3.42 \pm 0.10$  &  $4.7 \cdot 10^{-5}$  &  $1.1 \cdot 10^{-5}$
\\
triangular & 7.8    &  $1.8 \cdot 10^{-6}$   &  $3.21 \pm 0.09$  &  $4.9 \cdot 10^{-5}$  &  $1.1 \cdot 10^{-5}$
\\

uniform & 7.8  &  $1.7 \cdot 10^{-6}$  &  $3.18 \pm 0.08$  &  $4.8 \cdot 10^{-5}$  &  $1.1 \cdot 10^{-5}$
\\
\hline
Gaussian & 6    &  $4.3 \cdot 10^{-6}$  &  $2.96 \pm  0.09 $  &  $1.3 \cdot 10^{-4}$  &  $3.4 \cdot 10^{-5}$
\\
shifted& 6    &  $3.7 \cdot 10^{-6}$  &  $3.03  \pm    0.08 $  &  $1.1 \cdot 10^{-4}$  &  $2.9\cdot 10^{-5}$
\\
triangular & 6  &  $4.3 \cdot 10^{-6}$  &  $2.94  \pm 0.08$  &  $1.3 \cdot 10^{-4}$  &  $3.5 \cdot 10^{-5}$
\\
uniform & 6    &  $4.3 \cdot 10^{-6}$  &  $3.06  \pm    0.07$  &  $1.3 \cdot 10^{-4}$  &  $3.6 \cdot 10^{-5}$
\\

\hline
\end{tabular}
\caption{Summary of the power law fits to the success probabilities for various distributions ${\cal M}$,
with $Q$ varied over the range $[0.0,2.0]$, and for initial positions chosen randomly on $T^{1,1}$.
We fit the Monte Carlo data for $N_e>40$ to a power law of the form $P(N_e) = {\cal{A}}\,(N_e/60)^{-\alpha}$.}
\label{table:measure}
\end{center}
\end{table}





\subsection{An analytic explanation of the exponent $\alpha=3$}\label{analyticsection}

In our ensemble of potentials, inflation typically occurs near an approximate inflection point of the potential. We now show that a very simple model of single-field inflection point inflation, along the lines proposed in \cite{Freivogel:2005vv}, predicts $\alpha=3$, in excellent agreement with our numerical results.\footnote{We thank G. Shiu and H. Tye for very helpful discussions of this point.}

An approximate inflection point of a function $V(\phi)$ of a single field $\phi$ is a location where $V^{\prime\prime}=0$ and $V^{\prime}$ is small in appropriate units.  We choose the origin of $\phi$ to correspond to the zero of $V^{\prime\prime}$, so that
%
\begin{equation} \label{inflectionmodel}
V(\phi) = c_{0} + c_{1} \phi + c_{3} \phi^3 +
\ldots\,,
\end{equation}
with the $c_{i}$ being constants.  Assuming that the constant term dominates, the number of e-folds of inflation is
\begin{equation}
N_{e} \approx \frac{c_{0}}{\sqrt{c_{1}c_{3}}}
\end{equation}
in the regime of interest where the $c_{i}$ are small.

The approach suggested in \cite{Freivogel:2005vv} is to obtain the probability of $N_{e}$ e-folds of inflation by computing
\begin{equation} \label{unfixedpower}
P(N_{e}) = \int \prod_{i=1}^{k} d\xi_{i} F(\xi_{1}, \ldots \xi_{k})
\delta\Bigl(N_{e} - f(\xi_{1}, \ldots \xi_{k})\Bigr)\,,
\end{equation}
where the $\xi_{i}$ are the parameters of the model, $f$ is the number of e-folds as a function of these parameters, and $F$ is a measure on the parameter space.  Determining $F$ from first principles is very subtle, and is beyond the scope of this work. However, to compare to our numerical results involving {\it{relative}} probabilities of different numbers of e-folds, we need only use a measure $F$ that properly represents the measure ${\cal{M}}$ that we have imposed on the coefficients in our ensemble. At very small values of the $c_{i}$, we can approximate ${\cal{M}}$ as a constant, and so we take $F(c_{0},c_{1},c_{3}) =1$.  Thus, we need to evaluate
\begin{equation}
P(N_{e}) = \int d c_{1}\,  d c_{3}\, \delta\left(N_{e}
- \frac{c_{0}}{\sqrt{c_{1}c_{3}}}\right)\,,
\end{equation}
%
Performing the integral and again using the smallness of the $c_{i}$, we find
\begin{equation}
P(N_{e}) \approx -\frac{4c_{0}^{2}}{N_{e}^{3}}{\rm{log}}(c_0)
\end{equation}
so that $\alpha=3$,
which compares very well to our numerical results displayed in Table \ref{table:gamma}.

In the homogeneous background analysis described in \S\ref{probability}, the power in scalar perturbations is unconstrained. However, in \S\ref{experiment} we will assemble realizations whose scalar perturbations are consistent with the WMAP7 \cite{Larson:2010gs} normalization.  To compare to the ensemble of \S\ref{experiment} with fixed scalar power, we  must compute
\begin{equation} \label{fixedpower}
P(N_{e}) = \int \prod_{i=1}^{k} d\xi_{i} F(\xi_{1}, \ldots \xi_{k})
\delta\Bigl(N_{e} - f(\xi_{1}, \ldots \xi_{k})\Bigr)\delta\Bigl(A_{s}^{\star} - A_{s}(\xi_{1}, \ldots \xi_{k})\Bigr)\,,
\end{equation} where $A_{s}(\xi_{1}, \ldots \xi_{k})$ is the amplitude of the scalar perturbations as a function of the parameters $\xi_{i}$, and $A_{s}^{\star}$ is the central value measured by WMAP7.
For the inflection point model (\ref{inflectionmodel}),  when the   scalar power is fixed as in (\ref{fixedpower}), one again finds $\alpha = 3$, just as in the case (\ref{unfixedpower}) with unconstrained scalar power.
Moreover, the ensemble of \S\ref{experiment} with fixed scalar power is consistent with $\alpha=3$, providing a second check of our analytical model.

\subsection{Dependence on initial conditions}
\label{initial}

By construction, there is no preferred angular position selected by the ensemble of potentials:
upon averaging over all possible source locations in the bulk,
we   recover ensemble average rotational invariance.  However, in any particular realization, the potential will be quite different at different angular locations, so it is meaningful to ask about the effect of varying the initial angular position in a given realization.  Moreover, changes in the initial radial position can significantly alter the dynamics.  In \S\ref{inip} we determine the effects of altering the initial radial and angular positions, while the effects of varying the initial velocities are presented in \S\ref{iniv}.

\begin{figure}[!h]
  \begin{center}
    \includegraphics[width=4.0in,angle=0]{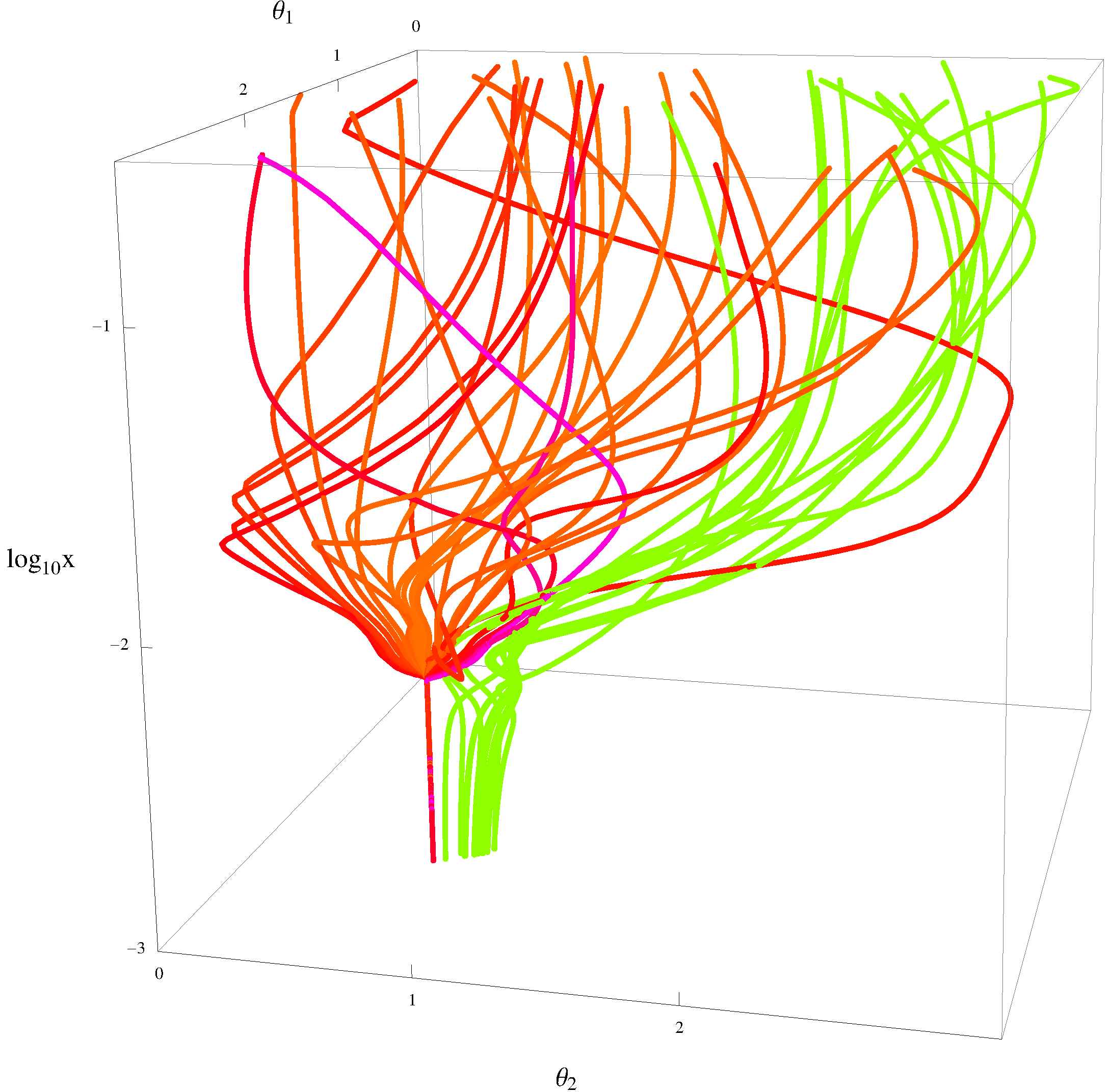}
    \caption{Trajectories in the $\theta_{1}-\theta_{2}$ plane, with $\rm{log}(x)$ vertical, for a fixed potential.  Notice the attractor behavior in the angular directions.  Green trajectories correspond to $\approx 5$ e-folds of expansion, while the remaining colors correspond to trajectories with $\approx 150$ e-folds.}
    \label{fig:attractor}
  \end{center}
\end{figure}

\subsubsection{Dependence on the initial position} \label{inip}

Prior works on initial conditions for  D-brane inflation have found that in many examples, the inflaton needs to begin  with small velocity just above the inflection point in order to yield substantial inflation.
In our ensemble, overshooting is not a problem: in most realizations yielding at least 60 e-folds, it suffices to begin the evolution with small velocity high up in the throat, e.g. at $x = 0.9$, while the inflection point is generally in the vicinity of $x=0.1$ or even smaller.  In fact, increasing the initial radial position  typically {\it{increases}} the amount of inflation. We suggest that this increase could be due to an increased opportunity to find the inflection point during a  prolonged period of radial infall.

The amelioration of the overshoot problem in our ensemble is a reflection of the difference between potentials that are fine-tuned by hand and potentials that are chosen randomly. In the former case, there is a natural tendency to fine-tune the potential to be just flat enough for 60 e-folds of inflation given perfect initial conditions, but no flatter.  In contrast, when  scanning through the space of possible potentials, one can actually find more robust examples.  As we have seen, successful realizations are reasonably common.

The success of inflation has very mild dependence on the initial angular positions: we find that in realizations of the potential that yield more than 60 e-folds of inflation for one set of initial positions,
an order-unity fraction of the space of initial angular positions leads to the same outcome.  Indeed, we find attractor behavior in the space of initial angles, as illustrated in Figure \ref{fig:attractor}.


\subsubsection{Dependence on the initial velocity}\label{iniv}

When the D3-brane begins with a radial velocity of order $10^{-6}$ of the local limiting speed (cf. \S\ref{dbi}), corresponding to an initial kinetic energy that is $\approx 10\%$ of the potential energy, it strikes the bottom of the throat within a fraction of an e-fold.  However, initial {\it{angular}} velocity of the same magnitude has a different effect: the D3-brane is quickly ejected from the throat.

We have found that two distinct causes contribute
to this ejection effect: first, D3-branes with large angular velocities can overcome potential barriers in the angular directions and thereby explore a larger fraction of $T^{1,1}$, including regions where the potential is strongly expulsive.  Second, angular momentum produces a barrier to radial infall, as in standard central force problems.  Inward-directed radial velocity and comparably large angular velocity have counterbalancing effects
in many cases, suggesting that slow roll inflation could arise in special regions of phase space where the initial velocities are not small, but have compensating effects.  We leave this as an interesting question for future work.

\subsection{The DBI effect}
\label{dbi}
In certain parameter regimes, higher-derivative contributions to the D3-brane kinetic energy can support a phase of DBI inflation \cite{Silverstein:2003hf,Alishahiha:2004eh}.  The
DBI Lagrangian is
\begin{equation}
{\mathcal L} = a^{3}\left( -T(y) \sqrt{1 - \frac{T_{3}g_{ij}\dot{y}^{i}\dot{y}^{j}}{T(y)}} - V(y) + T(y) \right),
\end{equation}
where in the $AdS_{5}\times T^{1,1}$ approximation with warp factor $e^{A}=x$, $T(y) = T_{3}x^{4}$.  DBI inflation can occur if the D3-brane velocity approaches the local limiting speed, i.e. if
\begin{equation}
 1 - \frac{T_{3}g_{ij}\dot{y}^{i}\dot{y}^{j}}{T(y)} \equiv \frac{1}{\gamma^2} \rightarrow 0 \,.
\end{equation}
In  our Monte Carlo trials,
we did not
observe a single example with  $\gamma-1 > 10^{-8}$, so the DBI effect was never relevant in our system.

To understand this result, we recall that steep potentials are generically required to accelerate D3-branes to approach the local speed of light, and not every potential that is too steep to support slow roll inflation is actually steep enough to drive DBI inflation in a given warped background.\footnote{We thank Enrico Pajer for instructive discussions of this issue.}  Specifically, DBI inflation requires \cite{Alishahiha:2004eh}
\bea \label{steep}
 \left(\frac{V'}{V}\right)^2 \gg \frac{T(y)}{V} \,,
\eea
so that for a fixed potential, DBI inflation could be achieved by appropriately reducing the background warp factor $T(y)$.  However, microphysical constraints prevent the warp factor from becoming arbitrarily small: when the infrared scale of a throat becomes small compared to the scale of supersymmetry breaking (due to e.g.\ fluxes or antibranes in a different region of the compactification), then relevant supersymmetry-breaking perturbations of the throat sourced in the ultraviolet lead to large corrections to the infrared geometry (cf.\ the discussion in \cite{Baumann:2010sx}).  This constraint enforces
\bea
	 V \lesssim   2T_{3} \,a_{0}^{4} \equiv 2 T(y)|_{\rm tip} \le 2 T(y) \,.
\eea
For comparison, the general arguments of \cite{Baumann:2008kq} concerning the scale of compactification corrections give $V_{\rm bulk} \sim T_{3} a_{0}^{4}$, and our choice to expand around $Q = 1$ is consistent with these results.
We have taken $T_3=10^{-2}$ and $a_{0} = 10^{-3}$ throughout, so the condition (\ref{steep}) could only be satisfied if a rather steep potential arose by chance.

In fact, an additional effect reduces the likelihood of DBI inflation in our analysis. For consistency we have restricted our numerical evolution to the region where the singular conifold approximation is applicable, which excludes the region of greatest warping, the tip of the deformed conifold.   In practice, we impose $x > 20\,a_{0}$, so that the minimum value of $T(y)$ explored in our simulations exceeds the global minimum value by a factor of $20^{4}$.  It would be very interesting to extend our analysis to the tip region, and as the unperturbed Klebanov-Strassler solution is well understood, it would be straightforward to incorporate purely kinematic corrections involving deviations of the field space metric from that of the singular conifold.  However, characterizing the structure of the potential in this region would be much more challenging, and would require understanding
the most general non-supersymmetric, perturbed solution for the tip region, along the lines of \cite{Baumann:2010sx} but departing from the approximately-conformal region.

One could also ask whether, even when the potential is not steep enough to accelerate the D3-brane to near the local speed of light, large initial velocities might still trigger a phase of DBI inflation.
We have found that cases with initial radial kinetic energy larger than 10\% of the initial potential energy typically strike the bottom without entering the DBI regime.
This result is compatible with prior investigations such as \cite{Bean:2007hc}, which found that exceptionally steep Coulomb potentials, which could arise in cones whose base spaces have extremely small angular volume, are required to produce DBI phases.

In summary, we find that the combination of the mild Coulomb potential in the Klebanov-Strassler throat, and general contributions to the D3-brane potential from moduli stabilization in the bulk, do not suffice to support DBI inflation in the $AdS_{5}\times T^{1,1}$ region.  It would be interesting to understand whether DBI inflation arises in the tip region \cite{Kecskemeti:2006cg,Pajer:2008uy}.

\section{Towards the Primordial Perturbations}\label{experiment}

D3-brane inflation generically involves at least six\footnote{In addition to the six D3-brane coordinates, the compactification moduli can also evolve.} dynamical fields, and the study of the primordial perturbations is rather intricate.  Most notably, entropy perturbations can be converted outside the horizon into curvature perturbations \cite{Gordon:2000hv}, provided that the inflaton trajectory bends in a suitable way.

In \S\ref{angularkinetic} we review the prospects for isocurvature-curvature conversion in warped D-brane inflation, following \cite{Chen:2008ada}.  We will find that although our setup provides an efficient framework for computing multifield effects, there is a wide range of parameter space in which these effects can be neglected. Therefore, in \S\ref{effectivelysinglefield} we restrict our attention to the subset of cases that admit a single-field description, and straightforwardly obtain the CMB observables.  A complete analysis of perturbations in the general case is postponed to a future publication.

\subsection{Angular kinetic energy and bending trajectories}\label{angularkinetic}
Typical trajectories that lead to prolonged inflation begin with relatively rapid angular and radial motion, then gradually spiral down to slow roll inflation along an inflection point, which is not necessarily parallel to the radial direction.  Eventually, the D3-brane leaves the inflection point and accelerates, ending slow roll; it then plummets towards the anti-D3-brane, triggering tachyon condensation and annihilation of the brane-antibrane pair.

\begin{figure}[!h]
  \begin{center}
    \includegraphics[width=3.16in,angle=0]{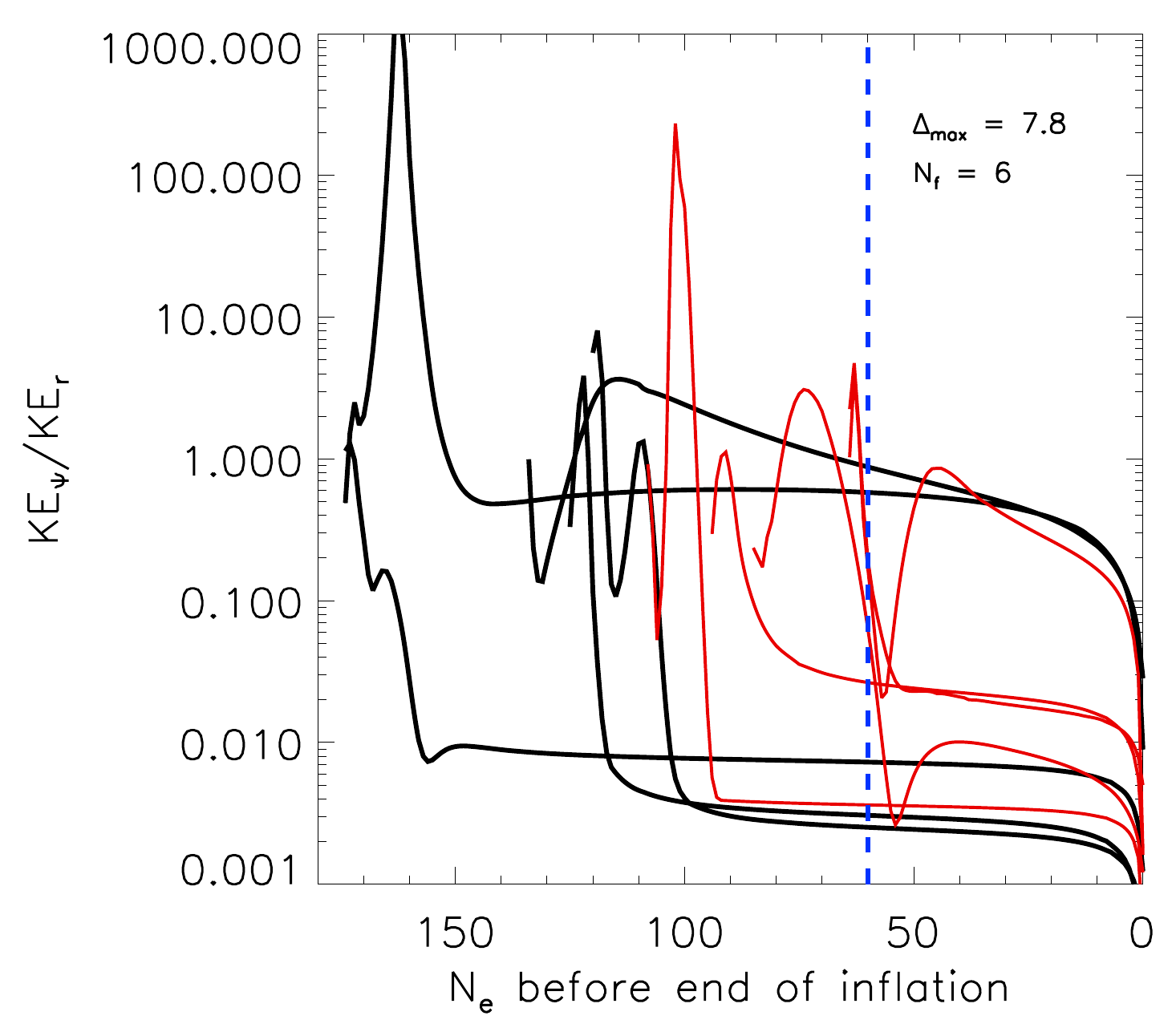}
     \includegraphics[width=3.0in,angle=0]{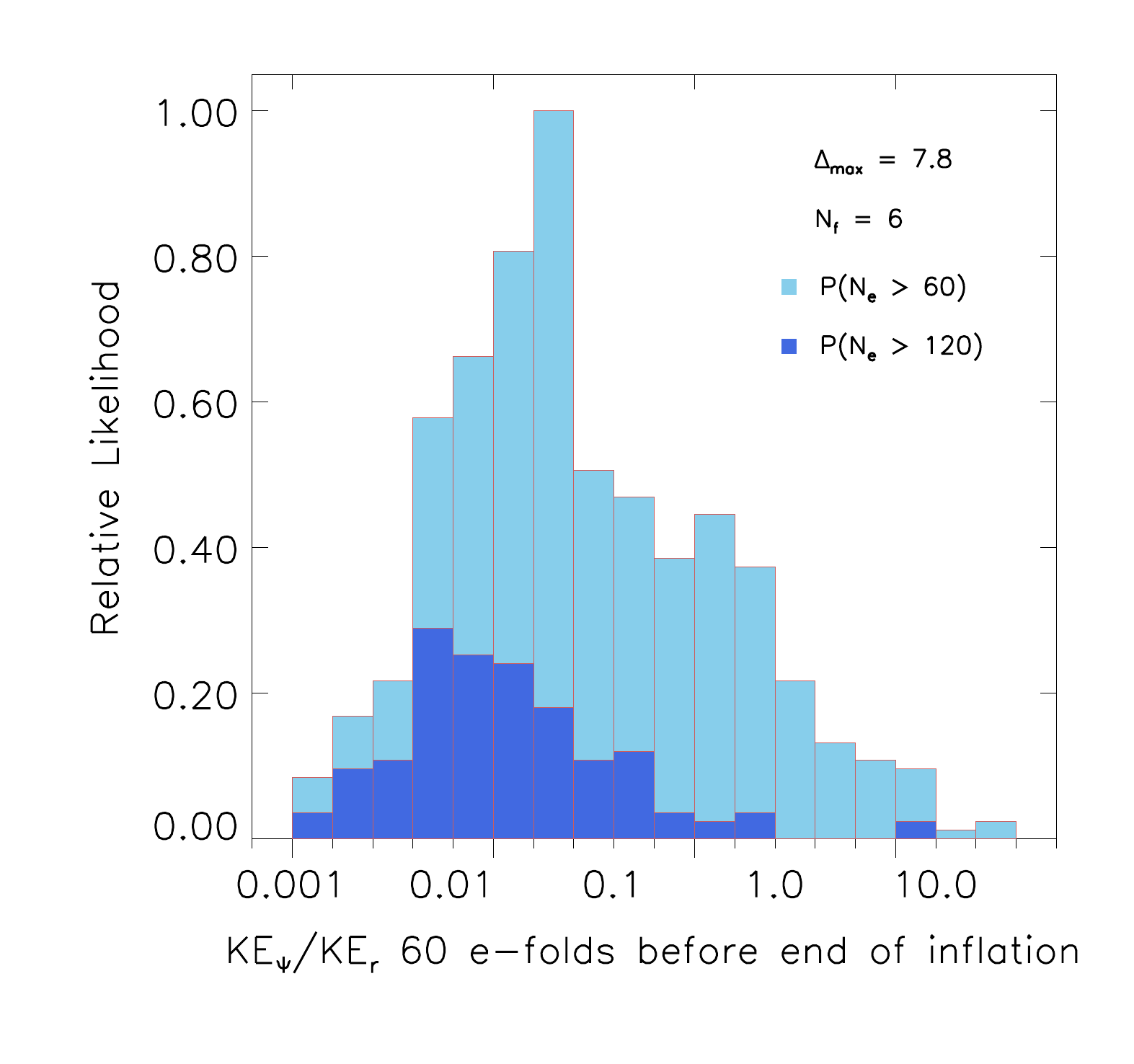}
    \caption{The ratio ${\rm{KE}}_{\Psi}/{\rm{KE}}_{r}$ of angular to radial kinetic energies for trials with $\Delta_{\rm max} = 7.8$ and $N_{f} = 6$.
      [Left panel]  Evolution of ${\rm{KE}}_{\Psi}/{\rm{KE}}_{r}$  for trials yielding $60 \le N_e \lesssim 120$ e-folds of inflation (red lines), and $120 \le N_{e} \lesssim 180$ e-folds (black lines).
     Notice that in some cases the angular kinetic energy is non-negligible, and nearly constant, in the final 60 e-folds, corresponding to an inflection point trajectory that is not purely radial.
     [Right panel]  Histogram of ${\rm{KE}}_{\Psi}/{\rm{KE}}_{r}$  60 e-folds before the end of inflation for potentials that yield more than 60 (light blue) or more than 120 (dark blue) e-folds of inflation.}
    \label{fig:KE}
  \end{center}
\end{figure}

We will begin by assessing the prevalence of multifield effects in our ensemble of realizations.
As a first, rough measure of the importance of multiple fields, one can examine the angular kinetic energy during inflation.
In Figure \ref{fig:KE}, we show the ratio of angular kinetic energy to radial kinetic energy as a function of 
the
number of e-folds before the end of inflation for selected examples.
In realizations yielding $N_{e}\approx 60$ e-folds, the ratio of angular kinetic energy to radial kinetic energy is often of order unity when observable scales exit the horizon, but diminishes thereafter.  In realizations yielding  $N_{e}\gtrsim 120$ e-folds, the transients are much diminished, but we still find cases (cf.\ Figure \ref{fig:KE}) in which the angular kinetic energy is of order the radial kinetic energy, and is approximately constant, throughout inflation.  These cases involve slow roll inflation along an inflection point that is not parallel to the radial direction.

Although some degree of bending is commonplace, we do not find  that the inflaton trajectory is substantially lengthened as a result of meandering \cite{Tye:2009ff} in six dimensions.
We find that the total distance $\ell$ in field space traversed in a realization with six active fields is negligibly larger than that for a realization with only one active field, $\ell(N_{f}=6) \lesssim 1.01 \ell(N_{f}=1)$.

Next, we turn to a more precise characterization of  multifield contributions to the primordial perturbations.  A comprehensive study of multifield effects in D-brane inflation \cite{Chen:2008ada} has been performed in the framework of \cite{Baumann:2007ah}, i.e.\ in terms of explicit embeddings of D7-branes in the Klebanov-Strassler solution.  One important lesson of \cite{Chen:2008ada} concerns the necessary conditions for isocurvature-curvature conversion at the end of inflation to make a significant contribution to CMB temperature anisotropies.  Under fairly general assumptions, this contribution is negligible unless slow roll persists into the deformed conifold region, and the Coulomb potential is subdominant to the moduli potential at the time of tachyon condensation \cite{Chen:2008ada}.
Our analysis applies only in the region above the tip of the deformed conifold, so we cannot consistently capture large multifield effects from the {\it{end}} of inflation.\footnote{As explained in \S\ref{dbi}, incorporating these effects would require an extension of the results of \cite{Baumann:2010sx} to the tip region, which is beyond the scope of this work.}

A further possibility is that a sharp bend in the trajectory partway through inflation will produce substantial isocurvature-curvature conversion and render invalid a single-field treatment of the perturbations.

To quantify the contributions from additional fields, we calculate $\eta$ in two components, $\eta^\parallel$ and $\eta^\perp$, as defined in \cite{GrootNibbelink:2001qt} and \cite{vanTent:2001sk}.
The acceleration of the inflaton parallel to its instantaneous trajectory is captured by $\eta^\parallel$, while  $\eta^\perp$ encodes the rate at which the inflaton trajectory  bends perpendicular to itself. Therefore, $\eta^\perp$ is an efficient measure of the role of multiple fields in producing the primordial perturbations  \cite{GrootNibbelink:2001qt}.
%
We define as ``effectively single-field" a realization in which the stringent cut $\eta^\perp/\eta^\parallel < 0.05$ is obeyed for the entirety of the last 60 e-folds.


Interestingly, we do find that in a small fraction of cases, abrupt angular motion occurs after a period of inflation driven by radial motion: the inflaton shifts rapidly from one angular minimum to another, then resumes radially-directed inflation.  These examples with $\eta^\perp/\eta^\parallel \gg 0.05$ require a full multifield treatment of the perturbations, and there is the intriguing possibility of substantial non-Gaussianity from superhorizon evolution of isocurvature perturbations.  We defer consideration of these interesting cases to a dedicated analysis \cite{forthcoming}.

\subsection{Single-field treatment of the perturbations}\label{effectivelysinglefield}

We now consider observational constraints on the substantial fraction of examples in which $\eta^\perp/\eta^\parallel < 0.05$, so that the primordial perturbations  are well-approximated by the single-field result.

We begin by computing the Hubble slow roll parameters,
\begin{equation}
\epsilon \equiv -\frac{\dot{H}}{H^2}=-\frac{d\log H}{d N_e}\,,
\end{equation}
\begin{equation}
\eta \equiv \epsilon-\frac{1}{2}\frac{d\log \epsilon}{d N_e}\,,
\end{equation}
We describe the power spectrum of curvature fluctuations using a normalization $A_s$ and scalar spectral index, or `tilt', $n_s$,  $\Delta_{{\mathcal R}}^{2}(k) = A_s (k/k_0)^{n_s-1}$.
In terms of $\eta_{60}$  and $\epsilon_{60}$, one has  $A_s=\frac{V_{60}}{24\pi^2\epsilon_{60}}$ and $n_s-1=2\eta_{60}-4\epsilon_{60}$, where the subscripts denote evaluation 60 e-folds before the end of 
inflation.

\begin{figure}[!h]
  \begin{center}
    \includegraphics[width=3.0in,angle=0]{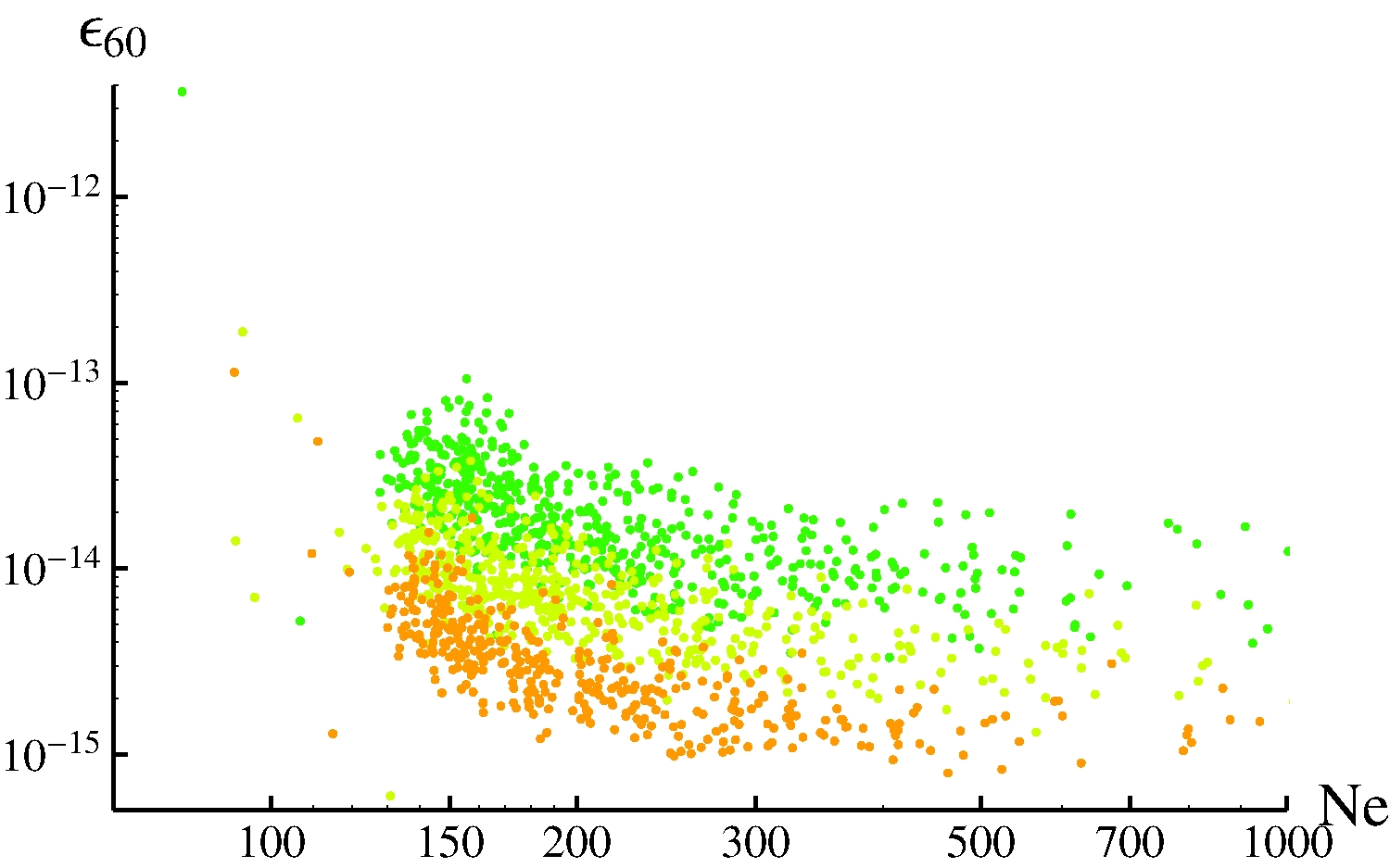}
    \includegraphics[width=3.0in,angle=0]{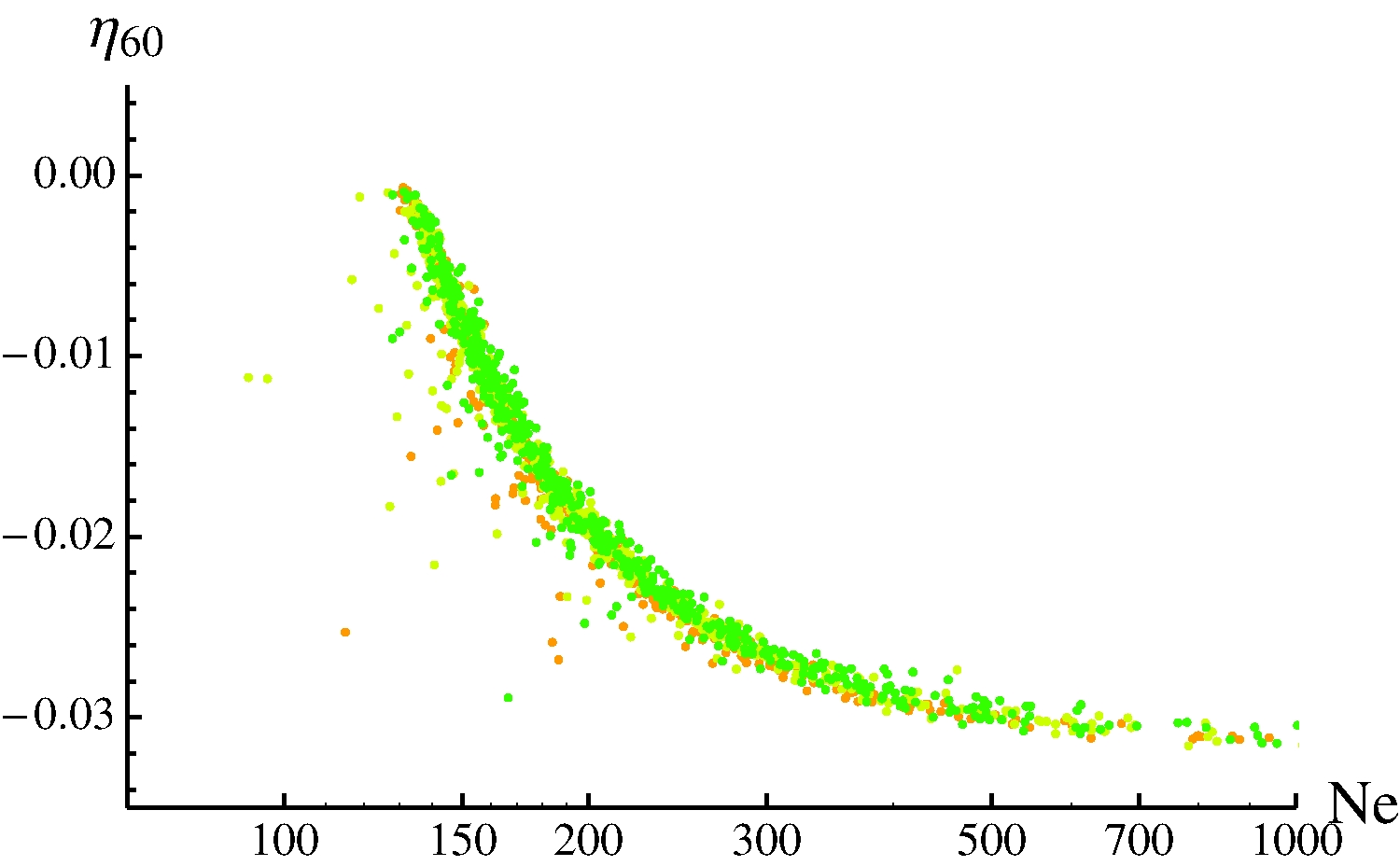}
    \caption{Hubble slow roll parameters $\epsilon_{60}$ and $\eta_{60}$ as a functions of the maximum number of e-folds. The scatter in $\epsilon_{60}$ for $N_{e} \gtrsim 120$ results from our scan over different values of the inflationary scale, as encoded in $a_0$.  Notice the absence of corresponding scatter in $\eta_{60}$.  Color coding: $-5 < \log_{10} a_0 < -4.75$, green; $-4.75 < \log_{10} a_0 < -4.5$, chartreuse; $-4.5 < \log_{10} a_0 < -4.25$, orange.  All points shown have trajectories with negligible bending, $\eta_{\perp} < .05 \eta_{\|}$.}
    \label{fig:epsandeta}
  \end{center}
\end{figure}

\begin{figure}[!h]
  \begin{center}
    \includegraphics[width=3.0in,angle=0]{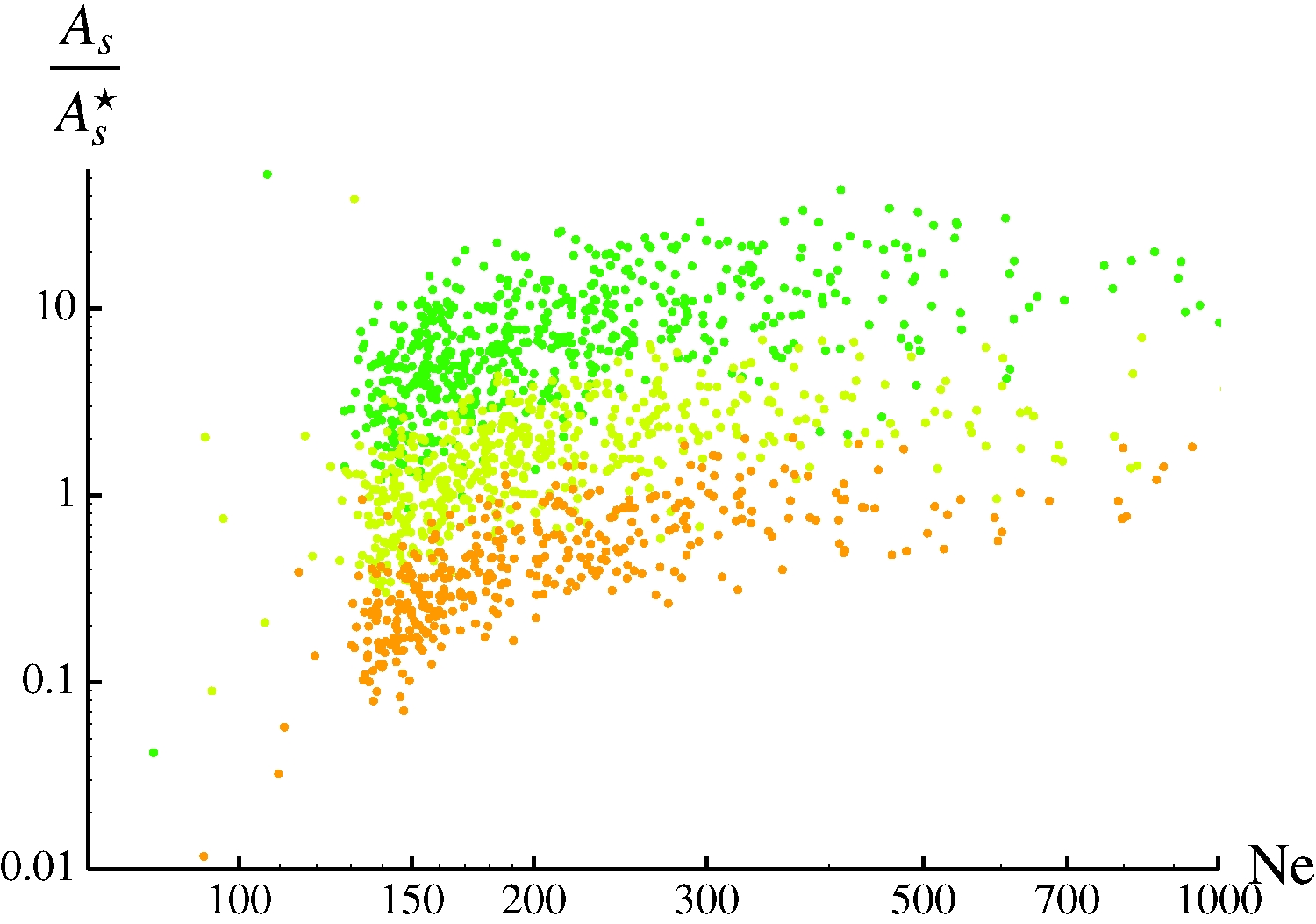}
\includegraphics[width=3.0in,angle=0]{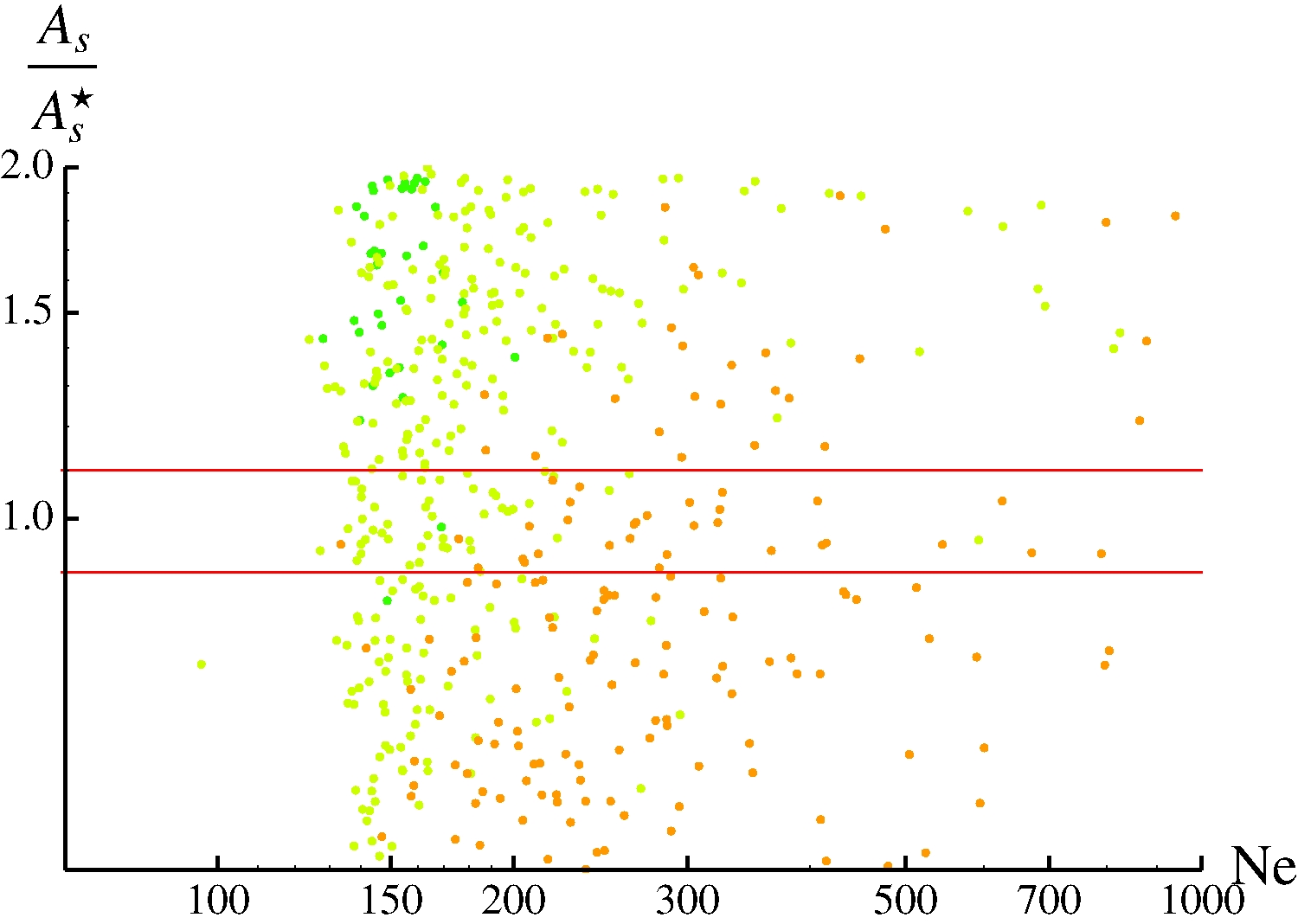}
    \caption{Scalar power $A_s$ compared to the measured central value, $A_{s}^{\star}=2.43\times 10^{-9}$, as a function of the total number of e-folds. On the right, we show a zoomed in version.  Within the red lines, the scalar power is consistent with WMAP7 at $2\sigma$.   All points shown have $\eta_{\perp} < .05 \eta_{\|}$, and the color coding is as in Figure \ref{fig:epsandeta}.}
    \label{fig:power}
  \end{center}
\end{figure}

\begin{figure}[!h]
  \begin{center}
    \includegraphics[width=4.5in,angle=0]{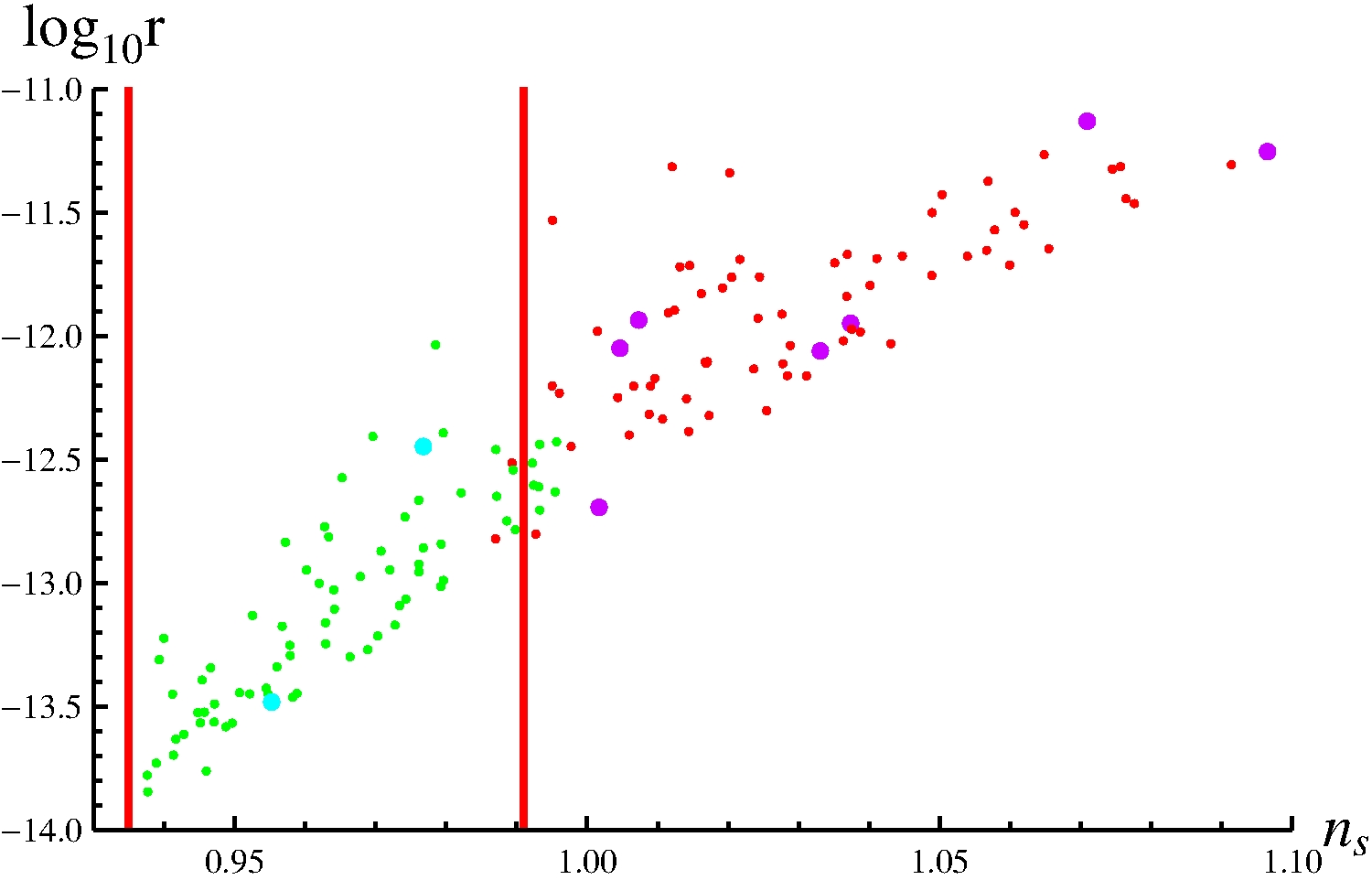}
    \caption{Scalar tilt $n_s$ and tensor-to-scalar ratio $r$ for cases for which the scalar power $A_s$ is consistent with WMAP7 at $2\sigma$.  The red lines indicate the region allowed at  $2\sigma$ by the WMAP7 constraints on $n_s$.   Green and red dots arise from realizations with $\Delta_{\rm max}=6$, while cyan and purple dots have $\Delta_{\rm max}=7.8$.  Green and cyan dots correspond to trajectories with negligible bending ($\eta_{\perp} < .05 \eta_{\|}$), while red and purple dots have $\eta_{\perp} \ge .05 \eta_{\|}$ and plausibly require a multifield analysis.  }
    \label{fig:rns}
  \end{center}
\end{figure}

The scalar power has significant dependence on the parameter  $D_{0}=2a_0^4 T_3$, which measures the height of the Coulomb potential.  We therefore scan over a range of values of $D_0$ (in practice, we fix  $T_3$ and scan over $a_{0}$), and for each successful trial that yields at least 60 e-folds, we compute the slow roll parameters, $A_{s}$, and $n_{s}$.
In Figure \ref{fig:epsandeta}, we show scatter plots of $\epsilon_{60}$ and $\eta_{60}$ as a function of the maximum number of e-folds.\footnote{Figures \ref{fig:epsandeta}, \ref{fig:power} and  \ref{fig:rns} share a set of $4.9 \cdot 10^{6}$ Monte Carlo trials at $\Delta_{\rm max}=6$,
out of which 8301 trials yield more than 60 e-folds and 140 also satisfy the WMAP7 constraints on $A_{s}$ at $2\sigma$. Figure \ref{fig:rns} additionally includes  $5 \cdot 10^{5}$ trials at $\Delta_{\rm max}=7.8$, out of which 750 examples yield more than 60 e-folds and 9 examples also satisfy the constraints on $A_{s}$ at $2\sigma$. All the data points given in Figures \ref{fig:epsandeta} and \ref{fig:power} obey $\eta^\perp/\eta^\parallel < 0.05$, while in
Figure \ref{fig:rns}, data points with $\eta^\perp/\eta^\parallel \ge 0.05$ are included, and indicated by red or purple  dots.}

Notice the paucity of examples with $N_{e} \lesssim 120$.  As $P(N_e) \propto N_e^{-3}$, a large fraction of trials yield $N_{e}$ in this range, but most such examples are excluded by the cut $\eta^\perp/\eta^\parallel < 0.05$.  This can be understood from Figure \ref{fig:KE}: realizations yielding $N_{e} \ll 120$ typically have substantial angular evolution in the final 60 e-folds, so that the single-field approximation is inapplicable.

\begin{figure}[!h]
  \begin{center}
    \includegraphics[width=3.5in,angle=0]{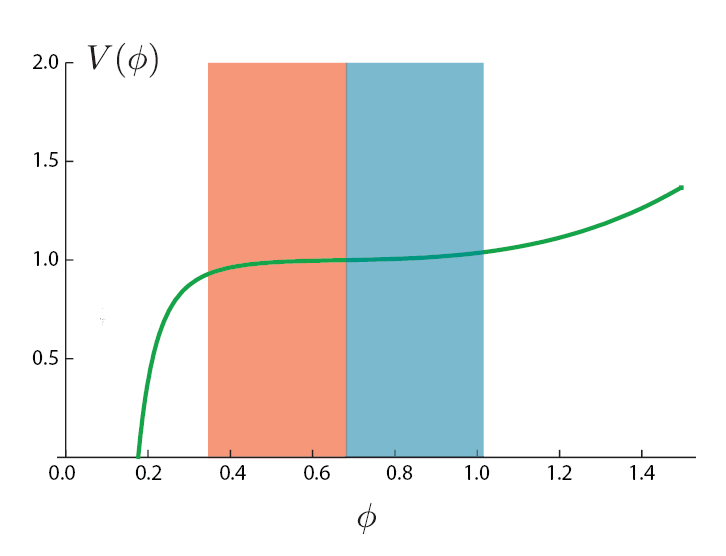}
    \caption{An inflection-point potential in one dimension, taken from \cite{Baumann:2007ah}.  The color coding indicates that if the inflaton is above the inflection point 60 e-folds before the end of inflation, $V^{\prime\prime}>0$ and the scalar power spectrum is blue. A red spectrum is possible if the inflaton has passed the inflection point 60 e-folds before the end of inflation.  }
    \label{fig:inflection}
  \end{center}
\end{figure}

We observe that $\eta_{60}$ is strongly correlated with the number of e-folds.
Importantly, when the single-field slow roll approximation is valid, only cases with $N_{e} \gtrsim 120$ e-folds are observationally consistent, since it is only for these cases that we have $V^{\prime\prime}<0$, ensuring $n_s<1$.  This is not surprising (cf.\ \cite{Baumann:2007ah}, Figure \ref{fig:inflection}): for single-field inflation in an approximate inflection point that is flat enough to yield exactly 60 e-folds of inflation, the CMB anisotropies are generated when the inflaton is above the inflection point, so that the potential is concave up, and hence $n_{s}>1$.
In a corresponding potential that yields 120 e-folds of inflation, observable modes  exit  the horizon when the inflaton is near to the inflection point (because 60 e-folds have elapsed and 60 e-folds  remain), so that for $\epsilon_{60} \ll 1$, one has $n_{s} \approx 1$.

The seven-year WMAP (WMAP7) constraints on $A_{s}$  and $n_{s}$  are $A_{s}=(2.43\pm0.11)\times 10^{-9}$  and $n_s=0.963\pm0.014$ at
$k=0.002$ Mpc$^{-1}$, at the 68\% confidence level \cite{Larson:2010gs}.
In Figure \ref{fig:power} we show a scatter plot of $A_s$ compared to the measured central value, $A_{s}^{\star}=2.43\times 10^{-9}$, indicating cases that are allowed at $2\sigma$ by the
WMAP7 constraint.
Notice that there is no fine-tuned choice of $D_0$, or of other input parameters, that guarantees a WMAP-normalized spectrum of perturbations: there is significant dispersion in $\epsilon_{60}$ in 
our ensemble of inflationary models, with corresponding dispersion in the scalar power.

Next, for the subset of cases  in which $A_{s}$ is consistent with experiment, we calculate the corresponding tilt $n_s$ and tensor-to-scalar ratio $r$, displaying the results in Figure \ref{fig:rns}.
Evidently, observational constraints on the tilt are readily satisfied for essentially all cases with $N_{e} \gtrsim 120$ e-folds, while cases with  $60 \lesssim N_{e} \lesssim 120$ e-folds solve the horizon and flatness problems but are not consistent with experiment.  We stress that this is a straightforward, albeit interesting, consequence of the inflection point form of the potential that is characteristic of successful realizations in our ensemble.

Within our model there is a microphysical upper bound $r \le r_{\rm max}$ on the tensor-to-scalar ratio resulting from the geometric bound on $\phi_{\rm UV}$  \cite{Baumann:2006cd}. However,
in single-field slow roll cases whose scalar perturbations are consistent with WMAP7, we find that $r \ll r_{\rm max}$.
It is sometimes argued that small values of $r$ require substantial fine-tuning.  Evaluating the absolute likelihood of inflation in this scenario requires information about a priori measures, and is beyond the scope of this work.  Even so, our analysis indicates that the requisite degree of fine-tuning is not extreme: in optimal regions of the parameter space, but without direct fine-tuning of the potential, we find examples consistent with all observations approximately once in $10^{5}$ trials, and it is clear from Figure \ref{fig:rns} that these scenarios have $r \lesssim 10^{-12}$.
\section{Conclusions} \label{conclusions}

We have performed a comprehensive  Monte Carlo analysis of D3-brane inflation for a general scalar potential on the conifold, obtaining robust predictions in spite of --- indeed, arguably because of --- the complexity of the inflaton action.  Our work builds on recent results \cite{Baumann:2010sx} that provide the structure of the potential, i.e.\ a list of possible terms in the potential with undetermined coefficients.  Most previous works (cf.\ e.g. \cite{Burgess:2006cb,Bean:2007hc,Baumann:2007np,Krause:2007jk,Baumann:2007ah,Ali:2010jx}) have treated special configurations in which suitably-aligned D7-branes stabilize the angular positions of the inflationary D3-brane, leading to one-field or two-field dynamics.  We have characterized the six-dimensional dynamics of the homogeneous background, without restricting the scalar potential.  We have not fine-tuned the potential by hand, but have instead drawn the coefficients in the scalar potential from suitable distributions, creating an ensemble of potentials, a subset of which led to inflation by chance.

We found that the probability $P(N_e)$ of $N_e$ e-folds of inflation is a power law, $P(N_{e}) \propto N_{e}^{-\alpha}$, with $\alpha \approx 3$. The exponent is robust against changes in our truncation of the potential, in the statistical distribution from which the coefficients in the potential  are drawn, in the initial conditions, and in the number of dynamical fields.
Moreover, we derived $\alpha=3$ from a simple analytical model of inflection point inflation.  This power-law behavior has significant implications for the prospect of detecting transients arising from the onset of inflation (cf.\ \cite{Freivogel:2005vv}): among all histories with at least 60 e-folds, histories with at least 65 e-folds --- in which most transients  are stretched to unobservable scales --- are considerably more likely.

When the inflaton starts at a radial location that is far above the inflection point, angular motion combined with gradual radial infall frequently allow the inflaton to reach the inflection point with a velocity small enough to permit inflation.
Moreover, we found attractor behavior in the angular directions: in an order-unity fraction of the space of initial angular positions, the inflaton spirals down to the inflection point.  However, large amounts of  radial or angular kinetic energy, of order the initial potential energy, are compatible with inflation only in exceptional cases.

DBI inflation did not arise by chance in our ensemble: the potential was never steep enough. It would be interesting to understand whether this finding can be generalized or is an artifact of the limitations of our treatment.

We have obtained the scalar perturbations for the subset of realizations in which a single-field description is applicable throughout the final 60 e-folds, deferring a comprehensive study of the multifield evolution of perturbations to future work.
In optimal regions of the parameter space, we found that 60 or more e-folds of inflation arose approximately once in $10^3$ trials, but because constraints on $A_s$ and $n_s$ enforce $N_e \gtrsim 120$, observational constraints were  satisfied approximately once in $10^5$ trials.  
Outside the optimal regions, the chance of inflation diminished rapidly.  As we lack a meaningful {\it{a priori}} measure on the space of parameters and initial conditions, we have not attempted to quantify the total  degree of fine tuning, but our results provide considerable information about relative likelihoods.

In the range of parameters where realizations consistent with observations of the scalar power spectrum are most likely, we found that the tensor-to-scalar ratio obeys $r \lesssim 10^{-12}$
in all examples allowed by WMAP7, which is much smaller than the maximum allowed by the Lyth bound.  
Our statements about the perturbations apply only to realizations that are consistently described by slow roll, effectively single-field inflation, which we checked by computing the rate of bending of the trajectory.  We anticipate that including more general multifield cases could populate additional regions of the $n_{s}-r$ plane.

Our findings have interesting implications beyond the setting of D-brane inflation.  The fact that our conclusions are unaffected by the statistical distribution used to generate the coefficients in the potential suggests that in inflationary models in which there are many competing terms in the scalar potential, the details of the individual terms 
can be less important than the collective structure.  A simplification of this form has been previously noted \cite{RMTNflation} in inflation driven by $D \gg 1$ fields with quadratic potentials \cite{Nflation}, where random matrix theory could be applied directly.{\footnote{See \cite{Frazer:2011tg} for recent related work.}}  Our present results suggest that this emergent simplicity may be  more general, and it would be valuable to understand whether a general inflationary model in a field space of dimension $D \gg 1$ has characteristic properties at large $D$.

\subsection*{Acknowledgements}
We are grateful to Thomas Bachlechner, Daniel Baumann, Richard Easther, Raphael Flauger, Ben Freivogel, Sohang Gandhi, Ben Heidenreich, Enrico Pajer, Gary Shiu, Henry Tye, Timm Wrase, and Jiajun Xu  for helpful discussions.   We particularly thank Dan Wohns for extensive discussions and suggestions, and Daniel Baumann and Raphael Flauger for their useful comments on the manuscript.  

N.A. and R.B.'s  research is supported by NSF CAREER grant AST0844825, NSF grant PHY0555216, NASA Astrophysics Theory Program grant NNX08AH27G and by Research Corporation.  The research of L.M. and G.X. was supported by the Alfred P. Sloan Foundation and by the NSF under grant PHY-0757868.  G.X. was supported in part by the Institute for Advanced Study, Hong Kong University of Science and Technology.  L.M. thanks the organizers of the Primordial Features and Non-Gaussianities workshop at HRI for providing a stimulating environment during the course of this work.

\appendix

\section{Structure of the Scalar Potential}\label{scalar potential}

In this appendix we summarize the computation of bulk contributions to the inflaton potential in warped D-brane inflation, following \cite{Baumann:2010sx}, to which we refer for notation and for further details.

D3-branes experience the potential
\begin{equation}
 V_{D3}=T_3\left(e^{4A}-\alpha \right)\equiv T_3\Phi_-\,.
\end{equation}
The classical equations of motion imply
\begin{equation}\label{fluxsquared}
 \nabla^2 \Phi_- =\frac{g_s}{96}|\Lambda|^2+\mathcal{R}_4 + {\cal{S}}_{{\sf
local}}\,,
\end{equation} where $\nabla^2$ is constructed using the unwarped metric on the compact space, $\Lambda$ is proportional to the imaginary anti-self-dual three-form flux, and ${\cal{S}}_{{\sf
local}}$ is a localized source due to anti-D3-branes.

The homogeneous solutions to (\ref{fluxsquared}) are harmonic functions on the conifold, i.e. solutions to the Laplace equation $\nabla^2 h=0$. Expanding $h$ in angular harmonics $Y_{LM}(\Psi)$ on $T^{1,1}$, we have
\begin{equation}
 h(x,\Psi)=\sum_{L,M}h_{LM}x^{\delta(L)}Y_{LM}(\Psi)+c.c.
\end{equation}
where $h_{LM}$ are coefficients, $L\equiv(j_1,j_2,R_f)$ and $M\equiv (m_1,m_2)$ label the $SU(2)\times SU(2)\times U(1)_R$ quantum numbers under the isometries of $T^{1,1}$, and the radial scaling dimensions $\delta(L)$ are related to the eigenvalues of the angular Laplacian,
\begin{equation}
 \delta(L)\equiv -2+\sqrt{H(j_1,j_2,R_f)+4}\,,
\end{equation}
where \cite{Ceresole}
\begin{equation}
H(j_1,j_2,R_f)\equiv 6(j_1(j_1+1)+j_2(j_2+1)-R_f^2/8)\,.
\end{equation}

Next, let us consider inhomogeneous contributions to the D3-brane potential sourced by imaginary anti-self-dual flux $\Lambda$.
Each mode of flux is a three-form with specified quantum numbers $L=(j_1,j_2,R_f)$ and $M=(m_1,m_2)$ under the angular isometries.
Thus, we can consider a general imaginary anti-self-dual flux $\Lambda$, compute\footnote{Some care is needed because certain contractions vanish identically.} $|\Lambda|^2$, and expand the result in terms of irreducible representations of the angular isometries. (That is, we expand products of two harmonic functions on $T^{1,1}$ in terms of individual harmonic functions, ensuring that each term still obeys the selection rules.)
Using the Green's function given in \cite{Klebanov:2007us}, one readily obtains the corresponding $\Phi_-$ profile.

The general potential for a D3-brane  therefore involves a sum of homogeneous terms $h$, inhomogeneous contributions sourced by flux, the Coulomb potential sourced by ${\cal{S}}_{{\sf
local}}$, and curvature contributions.  (For the computation of higher-order contributions from curvature, see \cite{Baumann:2010sx}.)



\end{document}